\documentclass[11pt]{article}

\usepackage[a4paper,hmarginratio=1:1,vmarginratio=2:3,
totalwidth=16.5cm,totalheight=21.2cm]{geometry}  
\usepackage{psfig}

\renewcommand{\baselinestretch}{1.28}

\newcommand{\be}{\begin{equation}}
\newcommand{\ee}{\end{equation}}
\newcommand{\bea}{\begin{eqnarray}}
\newcommand{\eea}{\end{eqnarray}}

\newcommand{\cE}{{\cal E}}
\newcommand{\cF}{{\cal F}}
\newcommand{\cG}{{\cal G}}
\newcommand{\cH}{{\cal H}}
\newcommand{\cI}{{\cal I}}
\newcommand{\cJ}{{\cal J}}
\newcommand{\cK}{{\cal K}}
\newcommand{\cL}{{\cal L}}

\newcommand{\cO}{{\cal O}}

\newcommand{\cR}{{\cal R}}

\newcommand{\bZ}{{\bf Z}}
\newcommand{\bR}{{\bf R}}

\newcommand{\ra}{\rightarrow}

\newcounter{oldcounter}

\begin{document}
\begin{flushright} 
DAMTP-2003-03.\\
\end{flushright} 
\vspace{0.7cm}
\begin{center} 
{\Large {\bf Wilson lines corrections to gauge couplings  
\\
\bigskip 
from a field theory approach.\\}}
\vspace{1.4cm} 
{\bf D.M. Ghilencea}
\\
\vspace{0.7cm} 
{\it D.A.M.T.P., C.M.S., University of Cambridge} \\
{\it Wilberforce Road, Cambridge, CB3 0WA, United Kingdom.}\\
\bigskip 
 \end{center}
\vspace{1cm} 
 \begin{center}
 {\bf Abstract}\\
\vspace{0.2cm}
\end{center}
Using an effective field theory approach, we
address the effects on the gauge couplings of one and two additional 
compact dimensions in the presence of a constant background 
(gauge) field. Such background fields are a generic presence in models 
with extra dimensions and can be employed for gauge 
symmetry  breaking mechanisms in the context of  4D N=1 supersymmetric 
models.  The structure of the ultraviolet (UV) and infrared (IR) 
divergences that the gauge couplings develop  in the presence of 
Wilson line vev's is investigated. One-loop radiative corrections to the 
gauge couplings due to overlapping effects of the compact dimensions 
and Wilson line vev's are computed for generic 4D N=1 models.
Values of Wilson lines vev's corresponding to
points (in the ``moduli'' space) of enhanced gauge symmetry cannot be
smoothly reached {\it perturbatively} from those corresponding 
to the broken phase.  The one-loop corrections are compared to
their (heterotic) string counterpart in the  ``field theory'' limit 
$\alpha'\ra 0$  to show remarkably  similar results when no
massless states are  present in a Kaluza-Klein tower. 
An additional correction to the gauge coupling
exists in the effective field theory approach  when for specific Wilson
lines vev's massless  Kaluza-Klein  states are present. This correction is
not recoverable by the limit $\alpha'\!\ra\! 0$ of the (infrared
regularised)  string because the infrared regularisation limit
and the limit $\alpha'\!\ra\!0$  of the string result do not commute.

\newpage\setcounter{page}{1}
$  $
\vspace{1.5cm}
\tableofcontents{}

%%%%%%%%%%%%%%%%%%%%%%%%%%%%%%%%%%%%%%%%%%%%%%%%%%

\newpage
\section{Introduction}

%%%%%%%%%%%%%%%%%%%%%%%%%%%%%%%%%%%%%%%%%%%%%%%%%%

 The  phenomenological and theoretical implications of the
physics of ``large'' extra  dimensions has recently  
attracted an increased research interest in the context 
of effective field theory approaches. String theory  
is ultimately thought to provide a complete and fully 
consistent  description of the  high energy physics.
Nevertheless,  effective field theory (EFT) approaches are able to 
describe accurately  many aspects of the physics of extra 
dimensions,  without  relying on  the string picture.

In this work we adopt an  effective field theory approach 
to  investigate the corrections to the gauge couplings in
4D N=1 supersymmetric  models with one and two extra dimensions,
in the presence of a constant background gauge field. 
The extra  dimensions may be compactified on manifolds 
or orbifolds thereof \cite{Dixon:jw} providing the possibility 
to construct chiral models. The compactification chosen  
and the value of the 
background gauge field (so-called Wilson lines' vev's 
\cite{Hosotani:1983xw}, \cite{Candelas:en}) 
have implications for  the one-loop  corrections to the 4D  gauge  
couplings and for the amount of gauge symmetry present. 
The interplay of these two effects on the 4D gauge couplings 
will be discussed in this work for the class of models considered. 
Let us first  present this problem in detail.

Additional compact  dimensions can  induce significant changes to 
the 4D gauge couplings. The one-loop corrected coupling 
in an  orbifold compactification to a 4D  N=1 supersymmetric model is
\begin{equation}
\frac{4\pi}{g^2_{i}(Q)} = \frac{4\pi}{g_{i}^{2}(M_s)}
+\frac{b_{i}}{2\pi } \ln
\frac{M_{s}}{Q}+\tilde\Omega_{i} +\cdots,
\qquad\quad \textrm{{\it i}: \, gauge \,\, group \,\, index.}
\label{gauge}
\end{equation}
$Q$ is a low energy scale above the supersymmetry breaking scale,
$M_s$ is the ultraviolet  scale or in the case of string theory, the 
string scale. In the EFT approach $g_i(M_s)$  
is the tree level (``bare'') coupling while in the (heterotic) string 
$g_i(M_s)$ is actually a   gauge  group independent
function of the so-called $S$ and $T$ moduli, invariant under
$SL(2,Z)_T\!\times \!SL(2,Z)_U\!\times \!Z_2^{T\leftrightarrow U}$
\cite{Nilles:1997vk}. This ensures that the string 
coupling is a well-defined expansion parameter, invariant 
under this symmetry of the string. Similar considerations may 
apply to other string models \cite{Antoniadis:1999ge}.
Further, the logarithmic  term in (\ref{gauge}) is due to 
the (infrared and ultraviolet regularised) contribution 
of the  light (``massless'') states charged under the 
gauge group. In a realistic  model these states are  N=1 multiplets and 
account for  the spectrum of the Minimal
Supersymmetric Standard Model (MSSM) or similar models.
In various compactifications to 4D such  states can 
arise as the 4D  Kaluza-Klein ``zero'' (or massless) modes of the 
initial (higher dimensional) fields  after compactification.

If the 4D N=1 string orbifold models  that we consider in this work 
(for a review see \cite{Bailin:nk}) 
have an N=2 sector of states (``bulk'') (e.g. $Z_4$ orbifold) other 
one-loop corrections ${\tilde \Omega_i}$ exist 
\cite{Dixon:1990pc}. Such orbifolds can also have N=4 sectors, but 
these do not affect the couplings due to the higher amount of supersymmetry.  
In an EFT description ${\tilde \Omega_i}$ accounts  for the sum of
individual one-loop corrections due to massive  Kaluza-Klein 
modes (non-zero levels) associated with the compactification 
and charged under the  gauge group. Such states which contribute  
are organised as N=2 multiplets
\cite{Kaplunovsky:1987rp,Dixon:1990pc}. 
In the heterotic string picture
${\tilde \Omega_i}$  includes  \cite{Dixon:1990pc} 
in addition the effect of the so-called winding states associated with
the extra dimensions and symmetries of the string (e.g. modular invariance)
and which have no EFT  description.  Other string 
constructions~\cite{Antoniadis:1999ge} bring tadpole cancellation 
constraints on  ${\tilde \Omega_i}$, again with no clear  EFT
equivalent, or may relate  ${\tilde \Omega_i}$  to the free energy of 
compactification \cite{Ferrara:1991uz}. Despite such additional string 
effects, the  EFT results and  the field theory  limit 
(i.e. infinite string scale) of some  string calculations 
can lead to  somewhat ``similar'' results for 
${\tilde \Omega_i}$. An example is the power-like dependence of the
couplings on the scale~\cite{Taylor:1988vt}.  However, understanding 
the  {\it exact} relationship between such approaches  requires a 
careful investigation.

A step towards clarifying this relationship is the analysis in 
\cite{Ghilencea:2002ff} where ${\tilde \Omega_i}$ was  
computed on pure EFT  grounds for 4D N=1
orbifold  compactifications with an N=2 sector 
(this is effectively a ``bulk'' as a two dimensional torus, while 
any completely untwisted N=4 sector of such orbifold does not 
affect $\tilde\Omega_i$). This calculation was done  by summing 
(infinitely many) one-loop  
corrections due to associated {\it massive} Kaluza-Klein  states.  
The result agrees with the limit 
$\alpha'\!\ra\! 0$ ($\alpha'\!\approx \! 1/M_s^2$)
of the  heterotic string  result \cite{Dixon:1990pc} due to  
{\it massive} Kaluza-Klein  and winding modes (in this limit string
effects such as winding modes effects were shown to be 
suppressed\footnote{Even in this 
limit winding modes  still play an UV role \cite{Ghilencea:2002ff} 
in  fixing the numerical coefficient of the 
UV leading term of  ${\tilde \Omega_i}$ when 
$\alpha'\ra 0$. At the EFT  level this coefficient is regulator 
dependent.}). However, differences emerge between  the EFT and string 
results \cite{Ghilencea:2002ak} when one includes  the
effect of the {\it massless} states of the theory (in this 
particular case these 
were  Kaluza-Klein states of level ``zero''\footnote{As we will
discuss, such massless states may also appear from  
{\it non-zero} Kaluza-Klein  levels if non-vanishing background fields 
(Wilson lines vev's) exist.}). The one-loop correction of the massless 
states is  infrared (IR) divergent both in  the EFT and in string
case. The differences mentioned  between the EFT and the limit 
$\alpha'\!\ra \!0$  of the string result
are  caused by the  infrared  regularisation of the string
 \cite{Ghilencea:2002ak}.  When the string IR  regulator is removed, 
this regularisation  discards $\alpha'$  dependent terms (divergent
for $\alpha'\!\ra\! 0$) which multiply the regulator. These terms  become 
relevant in the (field theory) limit $\alpha'\!\ra\! 0$ which does {\it not} 
commute with the IR regularisation of the string.   Such terms are
present in the final EFT correction to $g_i^2$
\cite{Ghilencea:2002ak}.  In the models we address
we will obtain such terms  whenever massless Kaluza-Klein 
modes are  present.

The discussion so far did not exhaust all possible one-loop effects 
in $\tilde\Omega_i$  that one encounters in string or EFT models with extra 
dimensions. Such models  usually have a larger amount of gauge
symmetry than the Standard Model does. A mechanism to reduce it  
is then required in a realistic model.
Introducing the usual Higgs mechanism is not always the most 
economical approach.  For multiply-connected manifolds a symmetry 
breaking  mechanism exists known as the Hosotani mechanism 
\cite{Hosotani:1983xw} or Wilson-line symmetry breaking
\cite{Candelas:en}.  
Explicit  models  of this type are known  at the string level  
\cite{{Ibanez:1986tp},{Ibanez:1987sn},{Ibanez:1987xa},{Ibanez:1987pj}}.
In this mechanism a constant  background  gauge field vev (of higher 
dimensional components of the gauge fields) controls 
the amount of gauge symmetry left after compactification. It also
affects the free energy of compactification \cite{LopesCardoso:1994ik} 
(also \cite{Ferrara:1991uz}) and may ``shift'' the  4D Kaluza-Klein
mass spectrum of the initial, higher dimensional fields. As a result 
of this change new radiative corrections  to the gauge couplings 
are expected.

The gauge symmetry breaking by Wilson lines is spontaneous and  
provides a  viable approach to model building (see  \cite{Hall:2001tn} 
for the relation to orbifold breaking). The effect of the Wilson
lines may be re-expressed as a ``twist'' 
in the boundary conditions for the initial 
fields (with respect to the compact dimensions) and which is removed 
by the limit of vanishing Wilson lines vev's. Since the breaking is 
spontaneous, the UV behaviour (encoded in the couplings $g^2_i$) of
the 4D models obtained after 
compactification should not be  worsened by non-zero Wilson lines
vev's. To check  this  we  evaluate  the 4D 
Kaluza-Klein masses  in the presence of  non-zero Wilson lines vev's, 
to  investigate   the radiative corrections   to $g^2_i$ and their 
dependence (continuity) on these vev's.  
We show that these corrections have a 
UV scale dependence similar to that when Wilson lines  have vanishing vev's.
As for the IR behaviour a regularisation is needed when for
{\it specific} Wilson line vev's some  Kaluza-Klein modes of non-zero 
level  may become massless and induce a gauge symmetry change.

Our intention  is  to present a general  EFT 
method  to compute radiative corrections to the 4D gauge couplings due to 
massive Kaluza-Klein states in the presence of Wilson lines.  
The framework is that of  4D N=1 supersymmetric models
with one and two compact dimensions, which correspond at a string level
to 4D N=1 orbifolds with an  N=2 sector of massive states,  
in the presence of Wilson line background \cite{Mayr:1995rx}.
The method evaluates 
the UV and IR behaviour of the couplings and their one-loop
correction in function of  the Wilson line vev's and may
easily be applied to specific models. 
Our results apply if the radii of compactification are 
``large'' (in units of UV cut-off),   without any reference 
to string theory. The EFT one-loop correction is compared to
its  (heterotic) string counterpart in the limit
$\alpha'\!\ra\! 0$ to find remarkably similar results. 
The correction may affect the unification of gauge couplings
in MSSM-like models derived from the heterotic string 
\cite{Nilles:1995kb}.
Wilson lines corrections to the gauge couplings were not 
computed  previously in a field theory approach. They were 
studied in  the heterotic string in \cite{LopesCardoso:1994ik},
\cite{Mayr:1995rx}. See  \cite{Friedmann:2002ty} for
compactification on $G_2$ manifolds.

The paper is organised as follows. In Section
\ref{wilson} we discuss the Wilson line breaking of 
the gauge symmetry and its effects on the  masses of 
4D Kaluza-Klein modes of the initial higher dimensional fields. 
We then address the effects of the Kaluza-Klein 
states and  Wilson lines vev's  on the gauge couplings (Section
\ref{wilson_c}). The  Conclusions are given in Section~\ref{conclusions}.  
The Appendix provides  extensive  technical  details for computing general 
Kaluza-Klein integrals used in the text (in DR and proper-time 
regularisation) and these results can  be used for 
other applications as well.

%%%%%%%%%%%%%%%%%%%%%%%%%%%%%%%%%%%%%%%%%%%%%%%%%%%%%%%%%%%%%%%

\section{Wilson line effects and extra dimensions.}\label{wilson}
\subsection{Definition of the models  and 
Wilson line symmetry breaking.}\label{wilson_symmetry}

%%%%%%%%%%%%%%%%%%%%%%%%%%%%%%%%%%%%%%%%%%%%%%%%%%%%%%%%%%%%%%%%

To begin with  we review the  gauge symmetry breaking  
by Wilson lines  and re-express it in terms of boundary conditions 
for higher dimensional fields. 
The class of EFT models considered in this work is  that of  4D N=1 
supersymmetric orbifold  models with gauge symmetry group $\cG$ 
larger than the  Standard Model (SM) group (e.g. SU(5)). The  models 
are assumed to have in addition to an N=1 spectrum 
an N=2 sector  of massive states (``bulk'') associated with 
the extra dimensions $y_m$ compactified on a circle ($m=1$) or a 
two-dimensional torus $T^2$ ($m=1,2$) and are regarded as the ``field
theory'' limit ($M_s\!\ra\!\infty$) 
 of a string compactification. A string embedding of
such models is  a compactification on $T^6/P$ with $T^6$ a
six-dimensional torus and $P$ the  point group, subgroup of SU(3)
\cite{Candelas3} (e.g. $Z_4$). 
The action of elements of $P$ on the six compact dimensions
can leave one complex plane ($y_m$)  unrotated (giving the N=2
sector),  rotate 
all three complex planes (the N=1 sector), and rotate none of them 
(the N=4 sector).  
Therefore the string compactification has in addition to the N=2
and   N=1 sectors,  an  N=4 sector as well. 
The N=1 sector gives the usual (MSSM-like) logarithmic
corrections to the gauge couplings while the  N=4 sector does not 
affect them. There remains the N=2 sector (``bulk'') of the unrotated  
plane ($y_m$) compactified on $T^2$ or a circle (if one dimension 
has radius set equal to $1/M_s$). Finally, a constant background (gauge)
field may be present. This gives the  string embedding  of our EFT models.
It also  justifies our considering of EFT models with extra dimensions 
$y_m$ compactified on a circle or a two-torus (corresponding to the N=2
sector) rather than on orbifolds thereof. From now on  we use an EFT 
approach and always refer to this sector only.  From the 4D perspective
towers of Kaluza-Klein  states  are present associated
with the compact dimension(s) $y_m$ and which correspond to 
initial, higher dimensional fields charged under $\cG$.
In the string picture  these  modes build up together with the
winding modes,   N=2 multiplets   of  4D N=1 orbifolds with 
non-zero Wilson line vev's \cite{LopesCardoso:1994ik},~\cite{Mayr:1995rx}.

First,  the group $\cG$ can be broken spontaneously  by those Wilson  
lines vev's which   ``survive'' (i.e. commute with) any  
orbifold action on the fields (see \cite{Haba:2002py} for
 examples at the EFT level). A Wilson  
line operator is defined as
\begin{equation}\label{www}
W_i=e^{i \int_{\gamma_i} dy_m  A_{y_m}^I T^I}, \quad I=1,\cdots,
\textrm{rk}\, \cG 
\end{equation}
with a sum over $I$ and $m$ understood;  $\gamma_i$ labels the  
contour(s) of integration over the $i^{th}$ compact dimension(s) (cycle). 
For two extra dimensions $A_{y_1}$ and 
$A_{y_2}$ should commute, otherwise the 4D effective action would 
contain terms  $Tr F^2\propto Tr[A_{y_1},A_{y_2}]^2$ from the 
field strength in 6D \cite{jp}. For phenomenological purposes, we
would like to avoid such terms,  thus 
$A_{y_1}$ and $A_{y_2}$ will lie in the Cartan sub-algebra of 
the Lie algebra of $\cG$ (of rank $\textrm{rk}\cG$).
In (\ref{www}) $T_I$ stands for a generator of this sub-algebra.  
The gauge symmetry left unbroken   by the Wilson lines vev's 
is that whose generators commute with all $W_i$.  Using the
commutators  in the Weyl-Cartan basis \cite{Slansky:yr}
\begin{equation}
[T_I, T_J]=0, \quad
[T_I, E_\alpha]=\alpha_I E_\alpha,\quad  I,J=1, \cdots,
\textrm{rk}\,\cG;
\quad \alpha=1+\textrm{rk}\,\cG,\cdots, \textrm{dim}\,\cG.
\end{equation}
one shows\footnote{we also use that
$\exp(i u_I T_I)\, E_\alpha\, \exp(-i u_I T_I)=E_\alpha \exp(i u_I \alpha_I)$
where a sum over I is understood.}
\begin{eqnarray}\label{symmetry}
\big[W_i, T_I \big] & = & 0,  \qquad \quad \,\, i=1,2.\nonumber\\
\big[W_i,  E_\alpha \big] & = & 0 \quad  \Leftrightarrow \quad  
\rho_{i,\alpha}\equiv - \frac{1}{2\pi}
%%%% 
\int_{\gamma_i} dy_m  \!<\!A_{y_m}^I\!>\! \alpha_I=0 
\quad  \textrm{mod}\, n, \,\,\,\, n\in \bZ; \,\, i=1,2.
\end{eqnarray}
where a  sum over $I$ and $m$ is understood;  $\alpha$ has
components $\alpha_I$, $I={\overline {1,\textrm{rk}\cG}}$  
and denotes the root associated with the generator $E_\alpha$.
The first relation in (\ref{symmetry}) 
shows that the rank (rk) of the group is not changed, 
while the second relation controls  the amount of symmetry
breaking,  through the vev's $<\!\! A_{y_m}^I\!\!>$, $I=
{\overline {1,\textrm{rk}\cG}}$ of the  Wilson lines
in various directions in the root space.  If the constraint in the rhs of 
the second line of eq.(\ref{symmetry}) is not respected, some 
gauge fields ``outside'' the Cartan sub-algebra  become 
massive\footnote{Examples of such fields would be 
for the  SU(5) case the so-called X, Y fields.} 
and the gauge  symmetry $\cG$ is  reduced\footnote{If the Wilson line 
is not in the Cartan  sub-algebra, the rank can also be reduced.}. 
We denote by $\cG^*$ this remaining symmetry, 
 generated by $T_I$ ($I=1,\cdots, \textrm{rk}\,  \cG$) and 
those $E_\alpha$ with  vanishing $\rho_{i,\alpha}$, $i=1,2$ (if no 
such $E_\alpha$ existed, then  $\cG$ would be broken to a product 
of $U(1)$'s). Additional (model dependent) constraints may apply to 
$\rho_{i,\alpha}$  in the string case  which depend on the
embedding of the point group in the gauge group $E_8\times E_8$. To
keep the EFT approach general we do not impose such constraints 
but keep $\rho_{i,\alpha}$ as parameters
throughout the calculation. In the final EFT result such constraints 
can then easily be implemented. Further  insight into the symmetry breaking  
$\cG\ra \cG^*$ is gained by  using  a $y$-dependent gauge transformation 
as we discuss separately for one and two extra dimensions.

%%%%%%%%%%%%%%%%%%%%%%%%%%%%%%%%%%%%%%%%%%%

\vspace{1.1cm}
\subsection{Effects on the 4D  Kaluza-Klein mass spectrum: 
One extra-dimension.}

%%%%%%%%%%%%%%%%%%%%%%%%%%%%%%%%%%%%%%%%%%%%

Consider the  5D fields $A_{\tilde\mu}(x,y)$ and $\Phi(x,y)$ 
in the adjoint and fundamental representations of $\cG$ respectively, with
$x\in M^4$ and index $\,\tilde\mu=\{\mu,y\}$, $\mu={\overline
{0,3}}$. One has 
\begin{equation}\label{ccc}
A_{\tilde\mu}(x,y+2\pi R)= Q\, A_{\tilde\mu}(x)\, 
Q^\dagger, \qquad  \Phi(x,y+2\pi R)=Q \,\Phi(x,y)
\end{equation}
where $R$  is the radius of the extra dimension $y$ and $Q$ is
some global transformation. For our purpose  one can
actually set $Q=1$ as the conclusions below will not depend on this.
In the following we assume $A_y$  constant (position independent) 
and attempt to  ``gauge away'' the field  $A_y$ using a $y$ dependent
transformation $U(y)$ (or $U(y) Q^{-1}$ if $Q$ is included).
 The new fields are
\begin{eqnarray}\label{cccc}
A'_y(x,y) &=& 0\qquad \textrm{ if }\qquad U(y)\equiv
e^{-i  y A_y\quad}\nonumber\\
A_\mu'(x,y) &= & U(y) A_\mu(x,y)\, U^{-1}(y)=
 A^I_\mu(x,y) T_I + A_\mu^\alpha(x,y)\, E_\alpha \,
e^{-i y A_y^I\alpha_I}\nonumber\\
\Phi'(x,y)&=& U(y) \,\, \Phi(x,y)
\end{eqnarray}
where a sum over $I$ and $\alpha$ is understood.
Since the generators $T_I, E_\alpha$ form a  linear
independent set, the second equation says that the fields $A_\mu^{I}$ 
in the Cartan sub-algebra do not ``feel'' the ``background'' field 
$A_y$ and are invariant under  $U(y)$. However, fields outside the
Cartan algebra ($A_\mu^\alpha$) are
transformed and the same applies to the field in the fundamental
representation. The initial condition eq.(\ref{ccc})  is  then changed
into $A'_y=0$ and
\begin{eqnarray}\label{twisted}
A_\mu^{' I}(x,y+2\pi R)&=&
A_\mu^{' I}(x,y)\nonumber\\
A_\mu^{' \alpha}(x,y+2\pi R)&=& 
e^{-i 2\pi R A_y^I \alpha_I}\, A_\mu^{' \alpha}(x,y) \nonumber\\
\Phi'_\lambda(x,y+2\pi R)& = & 
e^{-i  2\pi R A_y^I \lambda_I}\,\, \Phi'_\lambda(x,y)
\end{eqnarray}
where $\lambda_I$ are the weights (eigenvalues of $T_I$) and $\Phi_\lambda$
denotes a component of the multiplet $\Phi$.
The new fields $A_\mu^{' \alpha}(x,y)$ and 
 $\Phi'_\lambda(x,y)$ satisfy modified (``twisted'')
boundary conditions which  induce non-zero mass shifts for their 4D
Kaluza-Klein  modes.

 A solution to eq.(\ref{twisted}) is
\begin{eqnarray}\label{expansion}
A_\mu^{' \alpha}(x,y)
 &\sim & e^{-i  y A_y^I \alpha_I} \sum_{n\in \bZ} e^{i y n/R} 
 A^\alpha_{\mu,n}(x),\qquad \,\,\,\,
  M_{n}^2(\alpha) =(n- R\!<\! A_y^I \!>\!\alpha_I)^2/R^2\nonumber\\
 \Phi'_\lambda(x,y) & \sim & e^{-i  y A_y^I\lambda_I} 
 \sum_{n\in \bZ} e^{i y n/R} \phi_{n,\lambda}(x),
 \qquad \quad M_{n}^2(\lambda) =(n- R \!<\!A_y^I\!>\! \lambda_I)^2/R^2
\end{eqnarray}
where the fields depending on $x$ only 
are 4D Kaluza-Klein modes for vanishing background $A_y$ and
$\lambda,\alpha$ denote the weights/roots of corresponding fields.
Using the Klein-Gordon  
equation\footnote{$(D_\mu D^\mu +\partial_y\partial^y)
\Psi(x,y)=0$, with $\Psi$ to denote $A'$ or $\Phi'$ 5D fields.}
one finds the mass of the 4D Kaluza-Klein modes written 
in the rhs of (\ref{expansion}). The gauge symmetry $\cG$ is reduced
(to $\cG^*$) since the mass  of the zero-modes of some of the gauge fields
$A_\mu^{'\alpha}$ may become  non-zero, while for  
$A_\mu^{' I}$ they remain massless.  Kaluza-Klein 
levels of $A_\mu^{'\alpha}$ 
are shifted by (non-zero) $\rho_\alpha\equiv - R \!<\!A_y^I\!>\! \alpha_I$ 
which  equals  the value of  $\rho_\alpha$ introduced in
eq.(\ref{symmetry}). Note that the ``twist'' of the
boundary conditions in (\ref{twisted})
is removed by the formal 
limit of vanishing Wilson line vev's ($<\!A_y^I\!>\ra\! 0$).
Eq.(\ref{expansion}) is used in Section \ref{one_dim} to 
compute one-loop corrections to the 4D gauge coupling(s) of $\cG^*$.

%%%%%%%%%%%%%%%%%%%%%%%%%%%%%%%%%%%%%%%%%%%%%%%%%%%%%%%%%%%%

\vspace{0.4cm}
\subsection{Effects on the  4D 
Kaluza-Klein mass spectrum: Two extra dimensions.}

%%%%%%%%%%%%%%%%%%%%%%%%%%%%%%%%%%%%%%%%%%%%%%%%%%%%%%%%%%%%

\begin{figure}[t]
\centerline{\psfig{figure=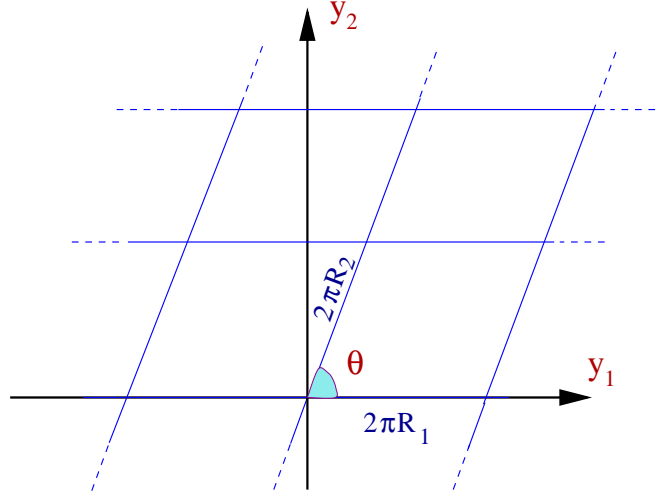,height=2.6in,width=3.4in,angle=0}}
\def\baselinestretch{1.1}
\caption{A two dimensional toroidal compactification  
constructed by identifying the opposite sites, with angle $\theta$ 
between the two cycles $\gamma_{1,2}$.
The points $(y1,y2)$ and $(y_1+2\pi R_1,y_2)$ are identified; 
the same applies to $(y_1,y_2)$ and
$(y_1+2\pi R_2 \cos\theta, y_2+2\pi R_2\sin\theta)$.
 $\gamma_{1,2}$ are defined along $R_1$ and $R_2$
directions respectively and $A_{y_{1,2}}$ along the orthogonal
dimensions $y_{1,2}$. 
}
\label{fig1}
\end{figure}

The  results of the previous section can be extended
to the case with two extra dimensions $y_{1,2}$, each compactified on
 a one-cycle $\gamma_i$, $i=1,2$ (Figure~\ref{fig1}).
Consider now the 6D fields $A_{\tilde\mu}(x,y_m)$ and $\Phi(x,y_m)$,
 in the adjoint  and fundamental representations 
of $\cG$ respectively, with  the index
$\tilde\mu~=~\mu,y_1,y_2$,  $x\in M^4$, $m=1,2$, ($\mu=\overline{0,3}$). 
Assuming constant $A_{y_{1,2}}$, one  
computes the Wilson lines $W_i$ of eq.(\ref{www})  corresponding 
to each $\gamma_i$. Using  definition (\ref{symmetry}) for 
$\rho_{i,\alpha}$  one can show 
\begin{equation}\label{wilsonvev}
\rho_{1,\alpha}=-R_1 \alpha_I <\! A_{y_1}^I\!>, 
\qquad \rho_{2,\alpha}=-
 R_2 \,\alpha_I 
\Big[\!\!<\!A_{y_1}^I\!> \cos\theta+<\! A_{y_2}^I\!> \sin\theta\Big]
\end{equation}
If {\it either} of the quantities in the rhs of (\ref{wilsonvev}) is 
non-integer/non-zero for some set of $\alpha$, the  symmetry $\cG$ is 
broken.  Note the dependence on the angle  $\theta$.

To compute the dependence of the 4D Kaluza-Klein masses on 
$\rho_{i,\alpha}$ we proceed as follows. First, 
one has the periodicity conditions (Figure \ref{fig1})
\begin{eqnarray}\label{periodic2D}
\Psi(x; y_1+2 \pi R_2 \cos\theta, y_2+2\pi R_2 \sin\theta)&=&
\Psi(x; y_1,y_2)\nonumber\\
\Psi(x; y_1+2 \pi R_1, y_2) &=& \Psi(x; y_1,y_2); \quad \Psi:
A_{\tilde\mu},\,\, or \,\, \Phi
\end{eqnarray}
where $\Psi$ stands for any of the  fields $A_{\tilde\mu}$ or $\Phi$.
These conditions are valid for an arbitrary two dimensional  toroidal
compactification (for an orthogonal  torus $\theta=\pi/2$ and  the
periodicity conditions on $(y_1,y_2)$ ``decouple'' to leave one such 
condition in each compact  direction).
As in the previous section, 
a $y$-dependent  transformation $V$ is introduced to ``gauge away'' the
fields $A_{y_1}$, $A_{y_2}$. 
One finds that the new (transformed) fields must satisfy 
\begin{eqnarray}\label{gauge2}
A'_{y_{1,2}}(x; y_1, y_2 ) &=& 0   
\qquad \textrm{if}\qquad V(y_1,y_2)=e^{-i  y_1 A_{y_1}-i  y_2 A_{y_2}
\quad}\nonumber\\
A'_\mu(x,y_1,y_2)&=& V A_\mu (x, y_1,y_2) \, V^{-1}=
 A_\mu^I(x,y_1,y_2) T_I +A_\mu^\alpha(x,y_1,y_2)\, E_\alpha \,
e^{- i (y_1 A_{y_1}^I + y_2 A_{y_2}^I) \alpha^I} \nonumber\\
\Phi'(x;y_1,y_2)&=& V\, \Phi(x; y_1,y_2)
\end{eqnarray}
with a summation over $I$ and $\alpha$ understood. With  $T_I$,
$E_\alpha$ linear independent one finds that the components $A_\mu^{I}$ 
are invariant under V,  $A_\mu^{'I}(x,y_1,y_2)=A_\mu^I(x,y_1,y_2)$ while
$A_\mu^\alpha(x,y_1,y_2)$ may not necessarily be so.
The initial periodicity conditions (\ref{periodic2D}) are changed into 
\begin{eqnarray}\label{twistedbc}
A_\mu^{' \alpha}(x; y_1+2\pi R_1,y_2)
&=& e^{{-i 2\pi R_1 A_{y_1}^I \,\alpha_I}}
\,\,A_\mu^{'\alpha}(x; y_1,y_2)\nonumber\\
A_\mu^{' \alpha}(x; y_1+2\pi R_2 \cos\theta,y_2+ 2 \pi R_2\sin\theta)
&=& e^{{-i 2\pi R_2 (A_{y_1}^I \cos\theta +A_{y_2}^I\sin\theta) \,\alpha_I}}
\,\,A_\mu^{'\alpha}(x; y_1,y_2)\nonumber\\
\Phi_\lambda'(x; y_1+2\pi R_2 \cos\theta,y_2+ 2 \pi R_2\sin\theta)
&=&e^{{-i 2\pi R_2 (A_{y_1}^I \cos\theta +A_{y_2}^I\sin\theta)\,
 \lambda_I}}\, \,
\Phi'_\lambda (x; y_1,y_2)\nonumber\\
\Phi'_\lambda (x; y_1+2\pi R_1,y_2)& =&
e^{{-i 2\pi R_1 A_{y_1}^I}\,\lambda_I} \,\, \Phi'_\lambda(x; y_1,y_2)
\end{eqnarray}
where $\Phi_\lambda$ denotes the component $\lambda$ of the 
multiplet  $\Phi$ and where we used that   
$T_I\,\Phi_\lambda\,=\,\lambda_I\,\Phi_\lambda\, .$
A solution to these equations has the structure
\begin{eqnarray}
\!\!\!\!\Psi'_\sigma(x; y_1, y_2)\!
&\sim& e^{-i  (y_1 A_{y_1}^I+ i y_2 A_{y_2}^I)\,\sigma_I}
 \sum_{n_{1,2}\in
\bZ} \psi_{n_1,n_2,\sigma} (x)\, u_{n_1,n_2}(y_1,y_2), 
\quad \textrm{with}\nonumber\\
\!\!\!\! u_{n_1,n_2}(y_1,y_2)\!\!&=&\!\!
\exp\bigg[ i \frac{n_1}{R_1} \left(y_1- \frac{y_2}{\tan\theta}\right)\!
+i  \frac{n_2}{R_2} \frac{y_2}{\sin\theta} \bigg], \quad 
\Psi'_\alpha\equiv\!  A_\mu^{'\alpha} \,\, and \, \, 
\Psi'_\lambda\equiv\!\Phi'_\lambda; 
\,\, \sigma=\alpha, \,\lambda
\end{eqnarray}
where the field $\Psi'$ is a notation for  either $A_\mu^{'\alpha}$ 
or $\Phi'_\lambda$ while $\sigma_I$ denotes  the roots $\alpha_I$ or the
weights $\lambda_I$, respectively. $u_{n_1,n_2}(y_1,y_2)$ is an 
eigenfunction of the Laplacian 
$\partial_{y_1}^2+\partial_{y_2}^2$ in the {\it absence} of 
the background gauge field because this was  already ``gauged away'' 
in (\ref{gauge2}) (to derive $u_{n_1,n_2}$ see for example
\cite{Dienes:2001wu}). 
If $\theta\!=\!\pi/2$, $u_{n_1,n_2}(y_1,y_2)$ 
is  a product of  one-dimensional eigenfunctions
 for each $y_i$.

Using the Klein-Gordon equation in 6D for massless fields
and  the above  mode expansion, one finds the mass of the 
Kaluza-Klein modes of 4D fields in the adjoint and fundamental representations
\begin{eqnarray}\label{mass2}
M^2_{n_1,n_2}(\sigma)& = & 
\left[\frac{n_1}{R_1}- A_{y_1}^I\sigma_I\right]^2
+\left[\frac{n_2}{R_2\sin\theta}-
\frac{n_1}{R_1 \tan\theta}-A_{y_2}^I \sigma_I\right]^2\nonumber\\
\nonumber\\
&\equiv & 
\frac{1}{\sin\theta^2 }\bigg\vert \, \frac{1}{R_2} (n_2+\rho_{2,\sigma})
- \frac{e^{i\theta}}{R_1} (n_1+\rho_{1,\sigma})\bigg\vert^2,
\quad \sigma=\alpha\,\,\, or \,\,\, \sigma=\lambda\label{mass2p}
\end{eqnarray}
It turns out that $\rho_{i,\sigma}$,  $i=1,2$
($\sigma=\alpha,\lambda$)  introduced in 
 the last step in eq.(\ref{mass2p}) have  
values equal to those found  in  eq.(\ref{wilsonvev})  using 
the definition of eq.(\ref{symmetry}). Here 
$\sigma\!=\!\alpha$ ($\sigma\!=\!\lambda$)  for the adjoint (fundamental)
representation. (we will  also use  the notation 
$v_\sigma\equiv\sigma_I\!<\!A_{y_1}^I\!>$,
$w_\sigma\equiv\sigma_I\!<\!A_{y_1}^I\!>$, (sum over $I$)).

If either $\rho_{1,\alpha}$, $\rho_{2,\alpha}$ are non-zero for a
fixed $\alpha$, there is no massless ``zero mode'' 
$(n_1,n_2)=(0,0)$ boson in the associated Kaluza-Klein tower. The initial 
symmetry $\cG$ is broken to a  sub-group $\cG^*$ generated by $T^I$ 
and those $E_\alpha$ for  which $A_\mu^{' \alpha}$ has a massless 4D 
zero-mode. This is controlled by the choice of $A_{y_i}^I$ vev's in
the root space of initial $\cG$. 
The scale where $\cG$ is broken to $\cG^*$ depends on the 
the potential developed by $A_y$ fields and is thus model dependent
(see \cite{Antoniadis:2001cv} for an example). Further, in specific cases
 $\rho_{1,\sigma}$ and $\rho_{2,\sigma}$ may be simultaneously 
integers for a fixed $\sigma=\alpha$ or $\lambda$ and
according to (\ref{mass2}) (if $\sigma=\alpha$)  there exists a
massless Kaluza-Klein boson  of non-zero level
$(n_1,n_2)\not\!=\!(0,0)$. Consequently a gauge symmetry  enhancement  
(beyond  $\cG^*$) takes place, enabled by additional corresponding 
$E_{\alpha}$.  These results agree with the previous findings in
Section \ref{wilson_symmetry}. 

One  may regard  the overall effect of symmetry breaking as a  
shift of   the Kaluza-Klein levels $n_i$ to ``effective'' levels  
$n_{i, eff}\!=\! n_i+\rho_{i,\alpha}$  of non-integer 
values. The shift is due to the geometry of compactification, 
eq.(\ref{periodic2D}) in the presence of
constant background fields whose  effect was replaced
by ``twisted'' boundary conditions, eq.(\ref{twistedbc}).
Eq.(\ref{mass2}) will be  used in Section \ref{twoextradim} to compute 
the radiative corrections to the 4D  gauge coupling(s) of the group $\cG^*$.

\section{Wilson line corrections to 4D gauge couplings.}\label{wilson_c}

For realistic  models  $\cG^*$ must include the Standard Model group  
(we will use $i=1,2,3$ to  label its component groups);  this can 
happen   if $\cG$ is SU(5). In general the  group $\cG$ may be larger
and $\cG^*$ contains additional group factors (not discussed). 
The  coupling $g_i$ of a group factor $i$ of
$\cG^*$ is 
\begin{equation}\label{formula}
\frac{4\pi}{g_i^{2}}\bigg\vert_{one-loop}=\frac{4\pi}{g_i^{2}}
\bigg\vert_{tree-level} +\Omega_i^T,\qquad \Omega_i^T\equiv \frac{1}{4\pi}
\sum_{\psi} \tilde\beta_{i}(\psi) \int_{0}^{\infty} \frac{dt}{t} \, e^{-\pi \, t
M^2_\psi/\mu^2}\bigg\vert_{reg.}
\end{equation}
$\Omega_i^T$ sums  all one-loop  corrections to the coupling $g_i$, 
induced by the states $\psi$  of mass $M_\psi$.
$\Omega_i^T$ is given by the  Coleman-Weinberg formal equation 
in the rhs of eq.(\ref{formula}) (for a discussion  see 
\cite{Kaplunovsky:1987rp}). In (\ref{formula}) $\mu^2$ is a 
{\it finite}, {\it non-zero} mass parameter introduced to 
enforce a dimensionless equation; the subscript 
``reg'' expresses that a regularisation of the integral is required.

$\Omega_i^T$ receives corrections  from the 4D massless and massive fields
charged under the group factor $i$ of $\cG^*$. For our purpose we  
need not specify the 4D massless or Kaluza-Klein zero
level spectrum, which  depends on further  details of the model 
considered. We restrict ourselves to  computing the structure of  
the corrections to $\Omega_i^T$  from  the 4D massive sector and  
we sum over all Kaluza-Klein
towers of states  charged under the group $i$ of $\cG^*$.
These~are: (1). 4D Kaluza-Klein towers of states whose levels
are not shifted  (i.e. $\rho_{k,\sigma}=0$, $k=1,2$) 
and have massless zero-modes. An example is (if $\sigma=\alpha$) 
that of Kaluza-Klein
states associated with the ``unbroken'' generators $E_\alpha$ 
(of the group $i$ of $\cG^*$) and:  
(2).   4D Kaluza-Klein towers of states of levels  shifted 
by the  amount $\rho_{k,\sigma}$, $k=1,2$. An example is
(if $\sigma=\alpha$)  that of
Kaluza-Klein states associated with the ``broken'' generators $E_\alpha$
(like $X,Y$ gauge bosons and their Kaluza-Klein tower 
for the SU(5) breaking to  the SM group).

For the beta functions $\beta_i$ one has (after suppressing the 
subscript $i$) that  $\tilde\beta(\sigma)=k_r
(\sigma_I \sigma^I)/{\textrm{rk}\cG^*}$ for $\sigma$ belonging  to 
representation $r$;  $k_r=\{-11/3,2/3,1/3\}$ for 
adjoint representations,  Weyl fermion and scalar respectively. 
The Dynkin index $T(r)= (\sum_{\sigma} \sigma_I\sigma^I)_r/(rk\cG^*)$ 
where the sum  is over all weights/roots $\sigma$
belonging to representation $r$, each occurring a number 
of times equal to its multiplicity \cite{Slansky:yr}.  
With the definition
$b_i(r)\equiv \sum_{\sigma}\tilde \beta_i(\sigma)$ for the weights
$\sigma$ belonging to $r$ one has
$b_i\!=\!-11/3\, T_i(A)~+~2/3 \, T_i(R)~+~1/3 \, T_i(S)$, to
account for the adjoint, Weyl fermion in representation $R$ 
and scalar in representation $S$. 
 Massive N=1 Kaluza-Klein states can be organised as
N=2 hypermultiplets with  $b_i=2 T_i(R)$ and  N=2 vector 
supermultiplets  with $b_i=-2 T_i(A)$.

\subsection{One extra-dimension and Wilson line corrections.}
\label{one_dim}

For the case of one additional compact dimension 
$\Omega_i^T$ can be written as
\begin{equation}\label{eq1}
\Omega_i^{T}=\sum_{r}\sum_{\sigma=\lambda,\alpha} 
\Omega_i(\sigma), 
\qquad\quad
\Omega_i(\sigma)\equiv \frac{1}{4\pi}
\sum_{m \in \bZ} \tilde\beta_i(\sigma) \int_{\xi}^{\infty}\frac{dt}{t} 
\, e^{-\pi\,t\, M^2_m(\sigma)/\mu^2} e^{-\pi \chi t}
\end{equation}
$\Omega_i(\sigma)$ is the contribution of a tower of
Kaluza-Klein modes  associated with a state $\sigma$  ``shifted'' by 
$\rho (\sigma)$ real, with  $\sigma=\lambda,\alpha$ the 
weight/root belonging to the representation r.
The sum over $m$ runs over all integers, representing Kaluza-Klein levels
of mass $M_{m}(\sigma)$ given by eq.(\ref{expansion})
\begin{equation}\label{massdef}
M_m(\sigma)=(m+\rho_\sigma)^2/R^2, 
\qquad \rho_\sigma= -R <\! A_y^I \!> \sigma_I\equiv  -R\, v_\sigma;\quad
\sigma=\alpha,\lambda.
\end{equation}
with $\sigma=\alpha$, $(\lambda)$ for the adjoint (fundamental) representation.
In eq.(\ref{eq1}) a regularisation of the integral was performed.
Since $\Omega_i^T$ is UV divergent ($t\ra\! 0$) an UV regulator 
$\xi\ra 0$ was introduced as the lower limit of the integral. 
For the special case when there are 
{\it massless} states in the Kaluza-Klein tower, the 
integral is also IR divergent $(t\ra\!\infty)$ and an IR  
regulator $\chi\ra 0$ is introduced.  
If no massless states exist in the Kaluza-Klein tower, one 
formally sets $\chi=0$. For other regularisations and 
their relationship with that employed here see Appendix A-4 of this work
and Appendix B, C  of \cite{Ghilencea:2002ff}.  From  
eq.(\ref{eq1}) the relation between the UV/IR regulators and their
associated mass scales can be inferred to  be of type 
$\Lambda^2\propto \mu^2/\xi$ for the  UV scale and  $Q^2\propto\chi
\mu^2$ for the IR scale.

As usually done  when computing the one-loop  corrections
in 4D compactified models \cite{Kaplunovsky:1987rp}, \cite{Dixon:1990pc}
we isolate in eq.(\ref{eq1}) the contribution of the
``zero'' modes (whose existence is model dependent, they may be
projected out by the initial orbifolding) 
from that of the non-zero level modes, which is general and is
computed in string case.  
To keep track of this separation we re-label by $\beta_i(\sigma)$ 
\,\, (${\overline \beta}_i(\sigma)$) 
the one loop beta function of ``zero'' (non-zero) modes, respectively. 
In the following the dependence  of 
$\rho$, $\overline\beta$, $\beta$ and $M_m$   on 
$\sigma=\alpha,\lambda$ is not  written explicitly. From (\ref{eq1}) we have 
\begin{equation}\label{omegai0}
\Omega_i  =  \frac{\beta_i}{4\pi} \cJ^0+\frac{\overline \beta_i}{4\pi}\cJ
\end{equation}
with the notation
\begin{eqnarray}
\!\!\!\!\!\!\!\!\
\cJ^0 & \equiv & \int_{\xi}^{\infty}\frac{dt}{t} 
\, e^{-\pi\,t \, M_0^2/\mu^2 } e^{-\pi \chi \,t}
=\Gamma[0,\pi\nu \xi (\rho^2+\chi/\nu)],\qquad\qquad
\nu\equiv \frac{1}{(R\mu)^2}
\nonumber\\
\nonumber\\
\!\!\!\!\!\cJ & \equiv & 
\sum_{m \in \bZ}' \int_{\xi}^{\infty}\frac{dt}{t} 
\, e^{-\pi\, t\, M_m^2/\mu^2} e^{-\pi \chi\,  t}\nonumber\\
\nonumber\\
 &=&
\frac{2 e^{-\pi\xi\chi}}{\sqrt{\nu\xi}}
-\Gamma[0,\pi\nu\xi(\rho^2+\chi/\nu)]+
2\pi ({\chi}/{\nu})^{\frac{1}{2}} 
\textrm{Erf}[\sqrt{\pi\xi\chi}]
-\ln\Big
\vert 2\sin\pi[\rho+i ({\chi}/{\nu})^{\frac{1}{2}}]\Big\vert^2
\label{omegai1}
\end{eqnarray}
with the functions $\Gamma[0,x]$ and Erf[$x$] defined in the Appendix
eq.(\ref{definition}).
A ``prime'' on the sum  over ``m'' indicates that the sum is
over all integers $m$ with $m\not=0$. To evaluate $\cJ$ the 
results of Appendix  \ref{r1ap}, eq.(\ref{R1}) were
used. Eq.(\ref{omegai1})  is valid if
\begin{eqnarray}\label{omegai2}
\nu \,\xi\ll 1,\qquad \textrm{or}\qquad\frac{1}{R^2}\ll
 \frac{\mu^2}{\xi}
\end{eqnarray}
Eqs.(\ref{eq1}) to (\ref{omegai2})
give  the most general result for the radiative correction to gauge
couplings. In the limit of ``removing'' the regulators dependence ($\xi\ra 0$) 
the $\Gamma$ functions contributions in $\cJ^0$, $\cJ$ can be
approximated by logarithms  while the Erf function
contribution vanishes.  The presence of the regulator  $\chi$  
 ensures that the result (\ref{omegai1})  applies whether or 
not there are massless states in the Kaluza-Klein
tower\footnote{Special care is needed when  removing the regulators
$\chi\ra 0$, $\xi\ra 0$ as these limits do  not always 
commute \cite{Ghilencea:2002ak}.}.

The mass parameter $\mu$ combines with the (dimensionless) regulators 
$\xi$ and $\chi$ to introduce the following associated mass scales 
\begin{equation} 
 Q^2\equiv \pi e^{\gamma} \chi\, \mu^2\bigg\vert_{\chi\ra 0}\, \qquad
  \Lambda^2\equiv \frac{\mu^2}{\xi}\bigg\vert_{\xi\ra 0}\
\end{equation}
$Q$ is therefore the low(est) energy scale, and $\Lambda$ 
is the high(est) energy (UV) scale of the theory.
With this notation the one-loop correction is 
 \begin{eqnarray}
 \!\!\!\!\!\!\!\!
 \rho=0:&&\!\!\Omega_i =\frac{\beta_i}{4\pi}\ln\frac{\Lambda^2}{Q^2}
  -\frac{\overline \beta_i}{4\pi}\ln\bigg[4\pi e^{-\gamma}
 (\Lambda R)^2\, e^{-2 \Lambda R}\bigg], \label{w1}\\
%%%%%
 \nonumber\\
 \!\!\!\!\!\!\!\!\rho\!\in\!\bZ^*:&&\!\! \Omega_i = 
 \frac{\beta_i}{4\pi}\ln\frac{\Lambda^2}{Q^2}
 -\frac{\overline \beta_i}{4\pi}\ln\bigg[4\pi e^{-\gamma}
 (\Lambda R)^2\, e^{-2 \Lambda R}\bigg]
 +\frac{{\overline{\beta_i}}-\beta_i}{4\pi} 
 \ln\bigg[1+\frac{ \pi e^{\gamma}(\rho/R)^2}{Q^2}\bigg]\label{w2}\\
\nonumber\\
%%%%%
 \!\!\!\!\!\!\!\!\!\!\!\!\!\!\!\!
 \rho\not\in\!\bZ: &&\!\!
  \Omega_i=\frac{\beta_i}{4\pi}\ln\frac{\Lambda^2}{Q^2}
 -\!\frac{\overline \beta_i}{4\pi}\ln \bigg[4\pi e^{-\gamma}
 (\Lambda R)^2\, e^{-2 \Lambda R}\bigg]
 - \!\frac{\overline \beta_i}{4\pi}\! 
 \ln\bigg[\frac{\sin(\pi \rho)}{\pi\rho}\bigg]^2\!\!\!\!
 -\frac{\beta_i}{4\pi}
 \ln\bigg[\frac{\!\pi e^{\gamma} (\rho/R)^2}{\!\!Q^2}\bigg]\label{w3}
\end{eqnarray}
where\footnote{We denoted by $\gamma$ the Euler constant, 
$\gamma=0.577216....$} 
according to eq.(\ref{massdef}), 
 $\rho/R$ is a  vacuum expectation value in a direction in the
weight/root space.
These equations are valid if the following conditions are respected:
\begin{eqnarray}
\rho=0: &&   Q^2\ll \frac{1}{R^2}\ll \Lambda^2\label{ww1}\\
\rho\in \bZ^*: && Q^2\ll \frac{1}{R^2}\ll \Lambda^2 
\quad \textrm{and} \quad\frac{\rho^2}{R^2}\ll \Lambda^2\label{ww2}\\
\rho\not\in \bZ: && \frac{1}{R^2}\ll \Lambda^2 \quad \textrm{and} \quad
\frac{\rho^2}{R^2}\ll \Lambda^2 \label{ww3}
\end{eqnarray}
Eqs.(\ref{w1}) to (\ref{ww3})  provide
the threshold  correction to the gauge couplings at one-loop level
due to a tower of Kaluza-Klein states in the presence/absence
of a constant background gauge field.

The first logarithmic term in eqs.(\ref{w1}), (\ref{w2}), (\ref{w3})  
stands for the contribution of the ``zero'' modes from the high scale
$\Lambda$ to some low  energy scale $Q$.  The second term in these equations
proportional to $\overline \beta_i$ contains  the linear  
contribution (in $\Lambda$) to the gauge couplings and is due to  the
Kaluza-Klein states of the extra-dimension, in the absence of Wilson
lines \cite{Ghilencea:2002ff}. Eq.(\ref{w1}) gives the correction
for the case of vanishing Wilson line vev's ($\rho=0$).

For the case $\rho\in \bZ^*$ the last term in eq.(\ref{w2}) 
gives an additional contribution  due to the Wilson line vev's. 
Note  that this correction is proportional to beta functions 
differences of zero and non-zero levels which may actually vanish in
specific cases. The correction
 also  depends on the low energy  scale $Q$ brought in by 
the need for an IR  regulator when   $\rho \in\bZ^*$.  
Consequently, the  low energy physics represented by $Q$  is related 
to Wilson line effects, even though the latter may take place at a
very high energy scale or may have a large vev  ($\rho$) compared to $Q$.
Formally, if one sets $\rho=0$ the case  of eq.(\ref{w1}) is recovered.
Finally, the  constraint (\ref{ww2}), $\rho/R\ll \Lambda$ may be 
``relaxed'' into $\rho/R\leq \Lambda$ or $\rho/R\approx \Lambda$ 
if one uses eq.(\ref{omegai0}), (\ref{omegai1}), (\ref{omegai2})
instead of (\ref{w2}), (\ref{ww2}).

The result for the  case with $\rho\not \in \bZ$ does not depend  on 
$Q$ (the dependence displayed in (\ref{w3}) cancels out). 
For $\Omega_i(\rho\!\not\in\!\bZ)$ the limit of reaching an integer 
or vanishing  $\rho$  is not finite, see  the last two terms in 
(\ref{w3}). Indeed, the one loop correction has  an infrared 
divergence at  integer  and vanishing values of  $\rho$, which 
are ``moduli'' points where new massless states appear. If these
correspond  to  vector superfields ($\sigma=\alpha$)
the symmetry    $\cG^*$ is enlarged. Therefore these  ``moduli'' 
points cannot be smoothly reached {\it perturbatively} by taking the 
limit of integer $\rho$.

We are now able to write the most general threshold correction to the
gauge couplings by combining the contributions $\Omega_i$ for 
all possible values for $\rho_\sigma$. 
Making the dependence on $\sigma$ manifest, one has from 
(\ref{eq1}) and  (\ref{w1}) to (\ref{w3})
\begin{eqnarray}
\Omega_i^T=\sum_{r}
\sum_{\sigma=\alpha,\lambda}
\Big[\Omega_i(\rho_\sigma=0)+ \Omega_i(\rho_\sigma\in\bZ^*)+
\Omega_i(\rho_\sigma\not\in\bZ)\Big]
\end{eqnarray}
with the remark that if the condition for the argument
of any term $\Omega_i$ is not respected, that term 
in the sum should be set to zero. Conditions
(\ref{ww1}) to (\ref{ww3}) should be considered accordingly.

\vspace{0.5cm}
\subsection{Two extra-dimensions and Wilson line corrections.}
\label{twoextradim}

The effects on the gauge couplings 
of the 4D Kaluza-Klein  states in the presence of
a  background (gauge) field  are similar to the one-dimensional
case. Eq.(\ref{formula}) becomes
\begin{equation}\label{QFTthresholds1}
\Omega_{i}^{T} \equiv \sum_{r} \sum_{\sigma=\alpha,\lambda} 
\Omega_i(\sigma),\qquad
\Omega_{i}(\sigma)\equiv \frac{1}{4\pi}
\sum_{m_{1,2} \in \bZ} \tilde\beta_{i}(\sigma) 
\int_{\xi}^{\infty}\frac{dt}{t} \, 
e^{-\pi\,t M^2_{m_1,m_2}(\sigma)/\mu^2 } \, e^{-\pi \chi \, t}.  
\end{equation}
with a summation over all weights/roots belonging to representation $r$.
As in  the one dimensional case an UV regulator $\xi$ ($\xi\!\ra\! 0$) 
is  introduced since the integral is divergent at $t\!\ra\! 0$ and the
associated UV scale is then proportional to $\mu^2/\xi$. For 
special  values of Kaluza-Klein levels $(m_1,m_2)$ and 
Wilson lines vev's (see eq.(\ref{mass2})),
$M_{m_1,m_2}$ may vanish, and the integral 
becomes infrared  divergent  ($t\!\ra \!\infty$).  
An IR regulator  $\chi$ is introduced $\chi\!\ra\! 0$  
for these particular cases, with associated infrared scale 
$Q^2\propto \mu^2 \chi$. The regulator  plays the role of a 6D mass term 
in the Klein-Gordon equation which shifts the value of $M_{m_1,m_2}$ 
in eq.(\ref{mass2}). 
The cases with  $M_{m_1,m_2}\!=\!0$ for Kaluza-Klein bosonic  states of
non-zero level are  important since they signal an
enlargement of the gauge symmetry $\cG^*$.
The  mass  of the Kaluza-Klein states given in (\ref{mass2}) can be
written as 
\begin{eqnarray}\label{kkmass}
M^2_{m_1,m_2}(\sigma)&=&
\frac{\mu^2}{T_2 U_2} \, \vert m_2+\rho_{2,\sigma}-U
(m_1+\rho_{1,\sigma})\vert^2, \quad \sigma=\alpha,\lambda.
\end{eqnarray}
with the notation 
\begin{equation}\label{moduli}
U\equiv U_1+i U_2=R_2/R_1\, e^{i\theta},\,\,\,\, (U_2>0);\, \qquad
{T} (\mu) \equiv i \,T_2 (\mu)= i \mu^2 R_1 R_2 \sin\theta\, .
\end{equation}
The structure of the mass formula (\ref{kkmass}) is very 
general and also applies to string compactifications.
At the string level $U$ and $T$  have correspondents in 
the so-called moduli fields related to the complex 
structure and (imaginary part of) the K\"ahler structure of 
the two-torus respectively\footnote{In the heterotic 
string $T$ is expressed in string units 
($\alpha'\propto 1/M_S^2$). To be exact, 
it is  $T_2^*=T_2/\xi$ that we later refer to and 
which is expressed in UV  
cut-off scale units that plays effectively the role
that $T_2$ does in string theory.}. The notation in 
eq.(\ref{moduli})  is  introduced to facilitate a
comparison with  the string results where this notation is 
standard.  $\rho_{1,2}$ are related to the Wilson 
line vev's, see eq.(\ref{wilsonvev}), (\ref{mass2p}). The mass
(\ref{kkmass})  equals that encountered in (heterotic) string
 compactification for zero winding modes. This is expected, 
since an effective  field theory  approach corresponds to the 
case of an infinite string scale when winding modes are infinitely 
heavy and their effects are suppressed.

To compute $\Omega_i^T$ of (\ref{QFTthresholds1}) we isolate 
the contribution of $(0,0)$ mode from that of the rest of
the Kaluza-Klein modes (this is allowed after the regularisation of the 
integral in (\ref{QFTthresholds1})).
This separation  is needed for two reasons. First, the exact 
final spectrum of zero level or  massless modes depends on  
further details of the models  considered; in particular some 
$(0,0)$   modes may not ``survive'' the initial orbifold projections. 
Second,  string calculations of $\Omega_i$  \cite{Dixon:1990pc}, 
\cite{Mayr:1995rx} that we want to compare with 
only compute the effects of the  non-zero  levels. 
To keep track of this mode separation we re-label $\tilde \beta_i$ by 
$\overline\beta_i$ ($\beta_i$)   for  the Kaluza-Klein levels 
$(m_1,m_2)\not=(0,0)$,  ($(m_1,m_2)=(0,0)$) respectively.

The analysis below considers separately Case 1 of 
vanishing Wilson lines vev's $\rho_{k,\sigma}=0$, $k=1,2$ 
($\sigma$ fixed) 
as a reference for when we evaluate Case 2 of non-vanishing Wilson vev's. 
Case 2 (B) when the Wilson line breaks the gauge symmetry is 
the most interesting for phenomenology.

\subsubsection{Case 1. Wilson line background absent.}

In this case  there are vanishing
vev's of the Wilson lines $\rho_{k,\sigma}=0$, $k=1,2$ for a
particular set of $\sigma=\alpha,\lambda$, with corresponding 
$E_\alpha$ as  unbroken generators. 
If this is true for {\it all} $\alpha$
there is  no breaking of the initial symmetry $\cG$. 
For simplicity we assume this is indeed the case, otherwise the
discussion refers to the unbroken part $\cG^*$ of $\cG$ only
and the effect of Kaluza-Klein towers ``not shifted'' by $\rho_{k,\alpha}$.
Therefore the  problem is  that  of one-loop 
corrections to the gauge couplings due to the N=2
sector (of a 4D N=1 orbifold) compactified on a two-torus 
(no Wilson lines vev's)
and well-known in the heterotic string \cite{Dixon:1990pc}.
At the effective field theory level this calculation was performed in 
\cite{Ghilencea:2002ff}, \cite{Ghilencea:2002ak},  reviewed  here 
for later reference.
In eq.(\ref{QFTthresholds1}) the mass of the Kaluza-Klein states 
does not depend on $\sigma$ and the sum of $\tilde\beta_i(\sigma)$
over $\alpha,\lambda$, $r$ (for the 
group~$\cG$!) may be performed before the  integral itself.
We denote this overall sum by $b_i$, $\overline b_i$ 
for  $(0,0)$ and for $(m_1,m_2)\not=(0,0)$ modes respectively.
%%%
The total correction $\Omega_i^T$ has  then the structure
\begin{equation}
\Omega_i^T=\frac{b_i}{4\pi} \cJ_0^{(1)}+\frac{\overline b_i}{4\pi}\cJ^{(1)}
\end{equation}
with 
\begin{eqnarray}
\cJ_0^{(1)}  & \equiv & \int_{\xi}^{\infty} 
\frac{dt}{t} \, e^{-\pi t M_{0,0}^2/\mu^2} \, e^{-\pi \chi t}
\,= \int_{\xi}^{\infty} 
\frac{dt}{t} e^{-\pi \chi t} =\Gamma[0,\pi \xi\chi]\nonumber\\
\nonumber\\
\cJ^{(1)} & \equiv &\sum_{m_{1,2}}'\int_{\xi}^{\infty} \frac{dt}{t} 
e^{-\pi \,t\, M_{m_1,m_2}^2/\mu^2} e^{-\pi \chi t}
=\sum_{m_{1,2}}'\int_{\xi}^{\infty} \frac{dt}{t}  
e^{ -\frac{\pi\, t}{T_2 U_2}\,  \vert m_2-U m_1\vert^2 -\pi \chi t}\nonumber\\
\nonumber\\
&=&
 -\ln \left[4 \pi e^{-\gamma}\, e^{-\frac{T_2}{\xi}}
\, \frac{T_2}{\xi} U_2 \left\vert \eta(U)\right\vert^4 
\right] + \pi \chi T_2  \ln\bigg[4 \pi e^{-\gamma} U_2
\frac{\xi}{T_2} \bigg]\label{no_w}
\end{eqnarray}
and with
\begin{equation}
\chi\ll
\min\left\{ {U_2}/{T_2}, {1}/{(T_2 U_2)}\right\}; \qquad
\max\left\{ {U_2}/{T_2}, {1}/({T_2 U_2})\right\}
\ll {1}/{\xi}\label{condition}
\end{equation}
A ``prime'' on a double sum stands for a sum over all integers
$(m_1,m_2)$ except the mode $(0,0)$ considered  by  $\cJ_0^{(1)}$.  
An infrared regulator was introduced in
$\Omega_i^T$ before its  splitting into $\cJ_0^{(1)}, \cJ^{(1)}$, 
because $M_{0,0}=0$ according to (\ref{kkmass}). 
The correction $\cJ^{(1)}$  was  evaluated in detail  
in ref.\cite{Ghilencea:2002ak}.   Condition (\ref{condition})  
ensures that higher order corrections  in the regulators $\chi$ 
and $\xi$ vanish in the limit of removing them, $\chi, \, \xi\ra 0$.
From (\ref{no_w}) one finds the overall 
correction to the gauge couplings due to the  towers of 
Kaluza-Klein states associated with two dimensions  
 (and vanishing  Wilson line vev's)
\begin{equation}\label{finalresult}
\Omega_i^T=\frac{b_i}{4 \pi}
\ln \frac{\Lambda^2}{Q^2}
- \frac{{\overline {b}}_i}{4 \pi} 
\ln\Big[4 \pi e^{-\gamma} \, e^{-{T_2^*}} \, {T_2^*} \, U_2
\, \vert \eta(U)\vert^4\Big]
+\frac{{\overline {b}}_i}{4 \pi}Q^2 R_1 R_2 \sin\theta \, e^{-\gamma}
\ln \bigg[4 \pi e^{-\gamma} \frac{U_2}{T_2^*}\bigg]
\end{equation}
We introduced the notation
\begin{equation}\label{cutoff}
T_2^*\equiv\left.\, \frac{T_2}{\xi}\, \right\vert_{\xi\ra 0}
= \Lambda^2  R_1 R_2 \sin\theta, \qquad \quad \textrm{and}
\qquad 
\left.\Lambda^2\equiv\frac{\mu^2}{\xi}\right\vert_{\xi\ra 0}, 
\qquad Q^2\equiv \pi e^{\gamma}
\, \mu^2 \chi\, \bigg\vert_{\chi\ra 0}
\end{equation}
which will be used in the remaining sections. Eq.(\ref{condition}) becomes
\begin{eqnarray}\label{conditions1}
Q \ll \min \left\{\frac{1}{R_1}, \frac{1}{R_2 \sin\theta}\right\};
 \qquad 
 \max \left\{ \frac{1}{R_1}, \frac{1}{R_2 \sin\theta}\right\}
\ll \Lambda
 \end{eqnarray}
which  implies $\Lambda^2 R_1 R_2 \sin\theta\gg 1$ i.e. a large
area of compactification (in UV scale units). 
Note that $(R_2\sin\theta)$ plays the role of an
``effective'' radius of the extra dimension.
Eq.(\ref{cutoff}) clarifies the link between the
UV/IR regulators and the mass scales $\Lambda$ and 
$Q$ which emerge as corresponding UV and IR cut-off scales
respectively.

The first term in $\Omega_i^T$ is the usual one-loop logarithmic
correction which accounts for the effects of $(0,0)$
modes, from the high scale $\Lambda$ to the low energy scale $Q$.
The second term  accounts for the effects of 
the  {\it massive} Kaluza Klein states associated with the two extra 
dimensions and shows the usual  power-like dependence \cite{Taylor:1988vt}
on the UV scale since $\Omega_i^T\sim T_2^*\sim \Lambda^2$. Its
structure can  be compared with that of the corresponding (heterotic) 
string result in the limit of an infinite string scale or $\alpha'\ra
0$,  when the additional effects of 
 the winding states of the string are minimised\footnote{Such
additional effects are (related to) world-sheet instanton effects, 
vanish if $\alpha'\!\ra\! 0$ and have no EFT description.}. 
The string result in this limit agrees with that of field theory.
For a detailed discussion on the second term in $\Omega_i^T$ and its
link with the heterotic string see  \cite{Ghilencea:2002ff}.

There remains the last term in $\Omega_i^T$ i.e. $Q^2 R_1 R_2
 \sin\theta \ln U_2/T_2^*\sim \chi\ln\xi$  whose origin  was analysed 
in detail in  \cite{Ghilencea:2002ak}. Here we review briefly its significance.
This correction arises from  the UV divergent contribution $\cJ^{(1)}$ of the
 massive momentum states  in the presence of the infrared regulator 
$\chi$. The latter was required by the  (IR divergent) contribution of the
massless mode $M_{0,0}$.  Therefore the term $\chi\ln\xi$  
establishes an (infrared) link between the massive and massless sectors.
The term $\chi\ln\xi$ or the last term in $\Omega_i^T$ that it
induces must be kept in the final correction to the gauge 
couplings because the limits of removing the regulators 
$\chi\ra 0$  and  $\xi\ra 0$ do not commute.

The presence of the last  term in $\Omega_i^T$ shows that even though 
Kaluza-Klein states may have a very large mass, of the  order of the
compactification  scales,  their {\it overall} contribution 
is still proportional to a much lower scale $Q$, where they may 
actually be decoupled. The contribution of this term may be large since
$U_2 \ll T_2^*$ (equivalently $1/R_1\ll \Lambda$). However the
coefficient in front may be small $Q^2 R_1 R_2 \sin\theta\ll 1$ for our
result to be accurate, see conditions in eq.(\ref{conditions1}).
We conclude that the overall effect of the infinite tower of
Kaluza-Klein states on the gauge couplings cannot be split into 
massless and massive modes only,  and a  combined effect of these
mass sectors through infrared effects is present.

The last term in $\Omega_i^T$  has  no equivalent at
the string level \cite{Ghilencea:2002ak}.  To understand why this is so, 
note that at the string level, the UV regulator 
$\xi\propto 1/\Lambda^2$  has a 
``correspondent'' in $\alpha'\propto 1/M_s^2$ with the limit
$\xi\ra 0$ to correspond to an infinite string scale or $\alpha'\!\ra\! 0$. 
String calculations of $\Omega_i^T$ require an infrared 
regularisation\footnote{For various infrared regularisations of the
string see the Appendix of \cite{Dixon:1990pc}, \cite{Foerger:1998kw} 
and \cite{Kiritsis:1994yv}.}, so $\chi$ also has a string infrared 
regulator  correspondent that we denote  $\epsilon$, with $\epsilon\ra 0$. 
Therefore, a string  equivalent of the last term in $\Omega_i^T$ would 
be $\epsilon \ln \alpha'$. Such term does appear in string
calculations during  the string  infrared regularisation 
(see \cite{Ghilencea:2002ak} and Appendix A of \cite{Foerger:1998kw}). 
However in string calculations $\alpha'\not = 0$  
and consequently $\epsilon \ln\alpha'\ra 0$ in the final, infrared
regularised  string result when $\epsilon\ra 0$.  
The {\it subsequent} ``field theory''  limit $\alpha'\ra 0$ of this
string result will then miss the last term in  $\Omega_i^T$.
The origin of this discrepancy is that in specific cases (such as 
two extra-dimensions) the infrared regularisation of the (world-sheet 
integral of the) string and its ``field theory'' limit $\alpha'\ra 0$
 do not commute.

The conclusion is that  the UV behaviour found using effective 
field theory methods for a 4D N=1 orbifold compactification with 
N=2 sub-sector (two-torus) is not necessarily  that of its (infrared 
regularised) string embedding in the limit $\alpha'\ra 0$. This issue 
is closely related to the infinite number  of states in the
Kaluza-Klein  tower that one sums over, which bring about a 
``non-decoupling'' of the  UV effects from the low energy ($Q$) sector. 
This concludes our review  of the radiative corrections 
to the  couplings for vanishing  Wilson  lines vev's.
For more details on this issue see \cite{Ghilencea:2002ak}.

\vspace{0.6cm}
\subsubsection{Case 2. Wilson lines background present. }

In this case there is a non-zero Wilson lines background
$\rho_{i,\sigma}$, $i=1,2$, $\sigma=\alpha,\lambda$, that 
can in some cases ($\sigma=\alpha$) 
affect the symmetry  $\cG^*$. We distinguish two general 
possibilities for $\Omega_i(\sigma)$ of eq.(\ref{QFTthresholds1})

\vspace{0.4cm}
\noindent
{\bf (A).} $\rho_{1, \sigma}\in \bZ$, $\rho_{2, \sigma} \in \bZ$ ($\sigma$
fixed),  and 

\noindent
{\bf (B).} $\rho_{1, \sigma}\not\in \bZ$ and $\rho_{2, \sigma}$ real (or 
 $\rho_{2, \sigma}\not\in \bZ$ and $\rho_{1,\sigma}$ real)

\vspace{0.5cm}
\noindent
{\bf Case 2 (A).  Computing  $\Omega_i$ for 
$\rho_{1, \sigma}\in \bZ$, $\rho_{2, \sigma} \in \bZ$.}
\vspace{0.3cm}

\noindent
In this case there exists a non-zero  Wilson lines vev,  
with $\rho_{i,\sigma}$, $i=1,2$ simultaneously integers,  
for a  fixed $\sigma=\alpha$ or $\lambda$.
Therefore for  a specific value of the Kaluza-Klein levels, 
$M_{m_1,m_2}(\sigma)$  vanishes and if this corresponds to a 
bosonic state $\alpha$, an enlargement of the gauge symmetry 
$\cG^*$ takes place.
When  $M_{m_1,m_2}(\sigma)$ vanishes  an infrared regulator $\chi$ is required
for the corresponding tower of Kaluza-Klein states which contributes
to $\Omega_i(\sigma)$. This is given by (below the $\sigma$ dependence
will not be written explicitly)  
\begin{equation}\label{eq38}
\Omega_i=\frac{\beta_i}{4\pi} \cJ_0^{(2)}
+\frac{\overline \beta_i}{4\pi}\cJ^{(2)}
\end{equation}
where we isolated the contribution $\cJ_0^{(2)}$ of the $(0,0)$ mode 
with beta function  $\beta_i$,  
from that  of non-zero modes $\cJ^{(2)}$ with beta function  
${\overline \beta_i}$ and
\begin{eqnarray} 
\cJ_0^{(2)} & \equiv &\int_{\xi}^{\infty} \frac{dt}{t} e^{-\pi\, t\,
M_{0,0}^2} \, e^{-\pi \chi t}=
 \int_{\xi}^{\infty} 
\frac{dt}{t} e^{-\frac{\pi t}{T_2 U_2}\, \vert \rho_2 - U \rho_1
\vert^2 -\pi \chi t} =\Gamma\Big[0,\pi \xi  \Big(\!
\vert \rho_2 -U \rho_1\vert^2/(T_2 U_2) +  \chi\Big)\!\Big]\nonumber\\
\nonumber\\
\cJ^{(2)}& \equiv & 
\sum_{m_{1,2}}'\int_{\xi}^{\infty} \frac{dt}{t} 
e^{-\pi \,t\, M_{m_1,m_2}^2/\mu^2}   \, e^{-\pi \chi t}=
 \sum_{m_{1,2}}'\int_{\xi}^{\infty} 
\frac{dt}{t}  
e^{ -\frac{\pi\,t}{T_2 U_2} \, \vert m_2+\rho_2-U (m_1+\rho_1) \vert^2 - \pi
\chi t}\nonumber\\
\nonumber\\
& = &\cJ^{(1)}+
\Gamma\Big[0,\pi \xi \chi\Big]-
\Gamma\Big[0,\pi \xi \Big(
\vert \rho_2 -U \rho_1\vert^2/(T_2 U_2) +  \chi\Big)\!\Big]\label{j2}
\end{eqnarray}
with
\begin{equation}                  
\chi\ll
\min\left\{ {U_2}/{T_2}, {1}/{(T_2 U_2)}\right\}
\leq \max\left\{ {U_2}/{T_2}, {1}/({T_2 U_2})\right\}
\ll {1}/{\xi}
\end{equation}
For a simple evaluation of   $\cJ^{(2)}$ one adds and subtracts 
under its  integral the exponential evaluated for $(m_1,m_2)=(0,0)$, 
then shifts the summation variables $m_{1,2}$ 
by the integers $\rho_{1,2}$ respectively and finally isolates from
the sum the ``new'' $(0,0)$  mode. 
One then  recovers an integral equal to $\cJ^{(1)}$  of eq.(\ref{no_w}) plus
two  additional terms, giving the result (\ref{j2}).

Using the notation introduced in eq.(\ref{cutoff}) we obtain from
(\ref{eq38}) and (\ref{j2})  the  result
\begin{eqnarray}
\Omega_i &=& \frac{\beta_i}{4\pi}\ln \frac{\Lambda^2}{Q^2}
- \frac{{\overline {\beta}}_i}{4 \pi} 
\ln\Big[4 \pi e^{-\gamma} \, e^{-{T_2^*}} \, {T_2^*} \, U_2
\, \vert \eta(U)\vert^4\Big]
+\frac{{\overline {\beta}}_i}{4 \pi}Q^2 R_1 R_2 \sin\theta \, e^{-\gamma}
\ln \bigg[4 \pi e^{-\gamma} \frac{U_2}{T_2^*}\bigg]\nonumber\\
\nonumber\\
&+&
\frac{{\overline \beta_i}- \beta_i}{4\pi} \ln \bigg[1+\frac{
\pi e^\gamma \vert \rho_2-U \rho_1\vert^2 }{ Q^2 (R_2 \sin\theta)^2}
\bigg]\label{case2w}
\end{eqnarray}
with 
\begin{eqnarray}\label{conditions2}
Q \ll \min \left\{ \frac{1}{R_1}, \frac{1}{R_2 \sin\theta}\right\}; 
\qquad 
 \max \left\{\frac{1}{R_1}, \frac{1}{R_2 \sin\theta}, 
\frac{\vert\rho_2 - U \rho_1\vert}{R_2 \sin\theta}\right\}
\ll \Lambda
 \end{eqnarray}
Except its last term, the result for $\Omega_i$ is similar 
to that for vanishing  Wilson lines discussed in Case~1. 
The first, second and third terms in 
(\ref{case2w}) account for the effects of the $(0,0)$ modes, massive 
Kaluza-Klein modes alone and  the ``mixed'' contribution proportional 
to $Q^2$, respectively. However, there exists an additional term 
proportional to $({\overline \beta_i}-\beta_i$),  not present in Case 1,
and this is due to the effects of the Wilson lines alone.
This term vanishes if one formally sets $\rho_1=\rho_2=0$ when Case~1
is recovered. Note the similarity of the last term in $\Omega_i$ 
to that of the last term in  (\ref{w2})  of the one
extra-dimension case. The Wilson line contribution 
can also be written as 
\begin{equation}\label{defvw}
\frac{{\overline \beta_i}-\beta_i}{4\pi} \ln\bigg[1+\frac{\pi e^\gamma
(v_\sigma^2+w_\sigma^2)}{Q^2}\bigg], \qquad
v_\sigma\equiv\sigma_I\!<\!A_{y_1}^I\!>,\quad
w_\sigma\equiv\sigma_I\!<\!A_{y_2}^I\!>,\quad \sigma=\alpha, \lambda.
\end{equation}
where $v_\sigma$ and $w_\sigma$ denote a vev in a direction in the 
root ($\alpha$) or weight ($\lambda$) space.  Note that  
$\beta_i$ and ${\overline \beta_i}$ depend on $\sigma$. This is
relevant when  the sum over $\sigma$ of eq.(\ref{QFTthresholds1}) is 
performed to compute total $\Omega_i^T$.

The correction due to Wilson lines vev's shows how the couplings 
 change when non-zero level Kaluza-Klein modes become
massless. The correction due to  the Wilson lines in eq.(\ref{case2w}) 
depends on their vev's and on the low scale $Q$, but
the ultraviolet  behaviour of the  models is not changed. 
Indeed the $T_2^*\sim \Lambda^2$  dependence of 
$\Omega_i$ is identical  to that of the case with 
vanishing Wilson lines vev's addressed in the previous section. 
Regarding the $Q$  dependence of $\Omega_i$ 
we remark the following.  
The  gauge couplings are changed at the {\it low} scale  $Q$ 
as a result of (possibly) {\it large} Wilson lines vev's $\rho_{i}$
and of their  near-cancellations against  large Kaluza-Klein 
levels $(m_1,m_2)$  corresponding to a  {\it large} momentum  in the 
compact directions $m_{1,2}/R_{1,2}$ and giving $M_{m_1,m_2}\approx 0$.

%%%%%%
\vspace{0.7cm}
\noindent{\bf Case 2 (B). Computing $\Omega_i$ for 
$\rho_{1,\sigma} \not \in \bZ$, $\rho_{2,\sigma}$ real (or 
 $\rho_{2,\sigma} \not \in \bZ$, $\rho_{1,\sigma}$ real).}
\vspace{0.4cm}

%%%%%%%%%%%%%%%%%%%%%%%%%%%%%%%%%%%%%%%%%%%%%%%%%%%%%%%%%%%

\noindent
In the following we assume $\rho_{1,\sigma} \!\not \in \!\bZ$ and
$\rho_{2,\sigma}$ real. (Appendix \ref{case3} extends this analysis 
to  $\rho_{2,\sigma} \not \in \bZ$, $\rho_{1,\sigma}$ real).
Phenomenologically this is the most interesting case we discuss.
There are non-zero Wilson line vev's and
$M_{m_1,m_2}(\sigma)\!\not=\!0$ for a fixed set of $\sigma=\lambda,\alpha$.
The corresponding generators $E_\alpha$ are broken, the
symmetry  is (reduced to) $\cG^*$ and Kaluza-Klein towers 
are ``shifted'' (by $\rho_{i,\sigma}$).
Since there  is no massless state 
for any integers $m_1,m_2$, in   eq.(\ref{QFTthresholds1})
there is no need for  an IR regulator
$\chi$ for the corresponding  correction $\Omega_i(\sigma)$ of the 
Kaluza-Klein tower. 
In the following the $\sigma$ dependence is not shown explicitly.
After  Poisson re-summation (\ref{p_resumation}) 
the integrand in  (\ref{QFTthresholds1})  becomes 
\begin{eqnarray}\label{sums}
\sum_{m_1,m_2}'e^{-\pi t M_{m_1,m_2}^2/\mu^2}
&=&\sum_{m_2}'e^{-\frac{\pi\, t}{T_2 U_2}  |m_2+\rho_2-U\rho_1|^2}
+\sum_{m_1}'\sum_{m_2}e^{-\frac{\pi\, t}{T_2 U_2} 
|m_2+\rho_2-U(m_1+\rho_1)|^2}\nonumber\\
\nonumber\\
&=&
  \sum_{m_2}' e^{-\frac{\pi\, t}{T_2 U_2}  |m_2+\rho_2-U \rho_1|^2}+
  \bigg[\frac{T_2 U_2}{t}\bigg]^{\frac{1}{2}}\sum_{m_1}'
e^{-\pi t \frac{U_2}{T_2}  \, (m_1+\rho_1)^2}\nonumber\\
\nonumber\\
&+&\bigg[\frac{T_2 U_2}{t}\bigg]^{\frac{1}{2}}
\sum_{m_1}'\sum_{\tilde m_2}' e^{-\pi {\tilde m_2}^2 \frac{T_2 U_2}{t} 
-\pi t \frac{U_2}{T_2}\, (m_1+\rho_1)^2+2 \pi i  {\tilde m_2} 
(\rho_2-U_1(\rho_1+m_1))}
\end{eqnarray} 
with a prime on the double sum in the lhs  to indicate 
the mode $(m_1,m_2)\!\not=\!(0,0)$ is not included.  Since $\rho_1$ is
non-integer  the above three contributions 
can be integrated separately over $(\xi,\infty$).

In $\Omega_i$ of (\ref{QFTthresholds1}) we isolate the contribution
of $(0,0)$  Kaluza-Klein levels ($\cJ^{(3)}_0$) 
from that of levels 
with $(m_1,m_2)\not=(0,0)$ and denoted $\cJ^{(3)}$,
with an obvious notation for the beta functions coefficients:
\begin{equation}\label{omega4}
\Omega_i=\frac{\beta_i}{4\pi} 
\cJ_0^{(3)}+\frac{\overline \beta_i}{4\pi}\cJ^{(3)}
\end{equation}
and
\begin{eqnarray}\label{j03}
\cJ_0^{(3)}&\equiv &\int_{\xi}^{\infty}\frac{dt}{t} e^{-\pi
t M^2_{0,0}/\mu^2}= \Gamma[0,\pi \xi \tau \vert\rho_2-U
\rho_1\vert^2], \qquad\quad \tau\equiv \frac{1}{T_2 U_2}\nonumber\\
\cJ^{(3)} &\equiv &\int_{\xi}^{\infty} \frac{dt}{t} 
\sum_{m_1,m_2}'e^{-\pi t M_{m_1,m_2}^2/\mu^2}
\equiv \cL_1+\cL_2+\cL_3,\qquad (m_1,m_2)\!\not=\!(0,0)
\end{eqnarray}
$\cL_1$, $\cL_2$ and $\cL_3$ denote the following integrals
of the three contributions given in the rhs of eq.(\ref{sums}) 
\begin{eqnarray}
\!\!\cL_1&\equiv &\int_{\xi}^{\infty}\frac{dt}{t}\sum_{m_2}'
e^{-\frac{\pi\, t}{T_2 U_2} \vert m_2+\rho_2-U\rho_1\vert^2}
%%%   =\cR_1(\tau\xi,\, \rho_2-U_1 \rho_1,\, U_2^2 \rho_1^2) 
 \nonumber\\
\nonumber\\
&=&\!\!\!
\frac{2}{\sqrt{\tau \xi}} \,  e^{-\pi \tau \xi \rho_1^2 U_2^2} 
+2\pi \vert \rho_1 \vert U_2\, \textrm{Erf}[\vert \rho_1\vert U_2
\sqrt{\pi\tau \xi}]
-\Gamma[0,\pi \tau \xi \vert \rho_2
-U\rho_1\vert^2]
-\ln\Big\vert 2 \sin\pi (\rho_2-U \rho_1)\Big\vert^2,
\nonumber\\
\nonumber\\
%%%%
\!\!\cL_2&\equiv & \!(T_2 U_2)^{\frac{1}{2}}\!
\int_{\xi}^{\infty}\frac{dt}{t^{3/2}}
\sum_{m_1}'e^{-\pi t\, \frac{U_2}{T_2}\,  (m_1+\rho_1)^2}
%%%% =\cR_2(\tau \xi U_2^2, \rho_1) \, \vert U_2\vert 
 \nonumber\\
\nonumber\\
&=& \!\!\!
-\frac{2}{\sqrt{\tau \xi}} \,  e^{-\pi \tau \xi \rho_1^2 U_2^2} 
-2\pi \vert \rho_1\vert U_2\, \textrm{Erf}[\vert \rho_1\vert U_2
\sqrt{\pi\tau \xi}]
+
\frac{1}{\tau \xi U_2}
+
2\pi  U_2 \Big[\vert \rho_1\vert + \frac{1}{6}+
\Delta_{\rho_1}^2-\Delta_{\rho_1}\Big],\nonumber\\
%%%%
\nonumber\\
\!\!\cL_3&\equiv &\!\!\! (T_2 U_2)^{\frac{1}{2}}\!
\int_{\xi}^{\infty}\!\frac{dt}{t^{3/2}} 
\sum_{m_1}'\sum_{\tilde m_2}' e^{-\pi \tilde m_2^2\frac{T_2 U_2}{t}-\pi t
\frac{U_2}{T_2} (m_1+\rho_1)^2+2 \pi i \tilde m_2 (\rho_2-U_1
(\rho_1+m_1))} 
\nonumber\\
\nonumber\\
&=&\ln \Big \vert 2 \sin \pi (\rho_2-U \rho_1)\Big\vert^2
-2\pi  U_2  \Big[\vert \rho_1\vert -
\Delta_{\rho_1}+\frac{1}{6}\Big]
-\ln\bigg\vert 
\frac{\vartheta_1(\Delta_{\rho_2}-U\Delta_{\rho_1}\vert U)}{\eta(U)}
\bigg\vert^2 
\label{cL123}
\end{eqnarray}
where  $\Delta_y$ denotes the  positive definite 
fractional part of $y$  defined as $y=[y]+\Delta_y$, 
$0<\Delta_y<1$, with  $[y]$ an integer number. 
$\vartheta_1(z\vert \tau)$ and $\eta(U)$ are special functions 
defined in the Appendix,  eqs.(\ref{R3_3}).

The above integrals are evaluated in detail in the 
Appendix, see  eqs.(\ref{R1}), (\ref{R2}) and (\ref{R3}) 
respectively. While evaluating them one introduces errors 
$\epsilon_i$ which are higher order corrections in the regulator 
$\xi$, see  eqs.(\ref{delta1}), (\ref{delta2}), (\ref{delta3}). 
Imposing that these corrections vanish gives the 
constraint
\begin{equation}\label{bound_xi}
{(T_2 U_2)}/{\xi} \gg
\max\Big\{ {1}/{U_2^2},\, U^2_2\Big\}
\end{equation}
which must be respected for eqs.(\ref{cL123}) to hold true.
Using the results of eqs.(\ref{omega4}), (\ref{j03}), (\ref{cL123})
and the notation introduced in eq.(\ref{cutoff}), we find the
result for the radiative correction to the gauge couplings
\begin{eqnarray}\label{omega_t} % \nonumber\\
\Omega_i & = & 
\frac{\beta_i}{4\pi}\ln \frac{\Lambda^2 (R_2 \sin\theta)^2}{\pi e^\gamma
\vert \rho_2 - U \rho_1\vert^2}
- \frac{{\overline \beta}_i}{4 \pi} 
\ln\Big[4 \pi e^{-\gamma} \, e^{-{T_2^*}} \, {T_2^*} \, U_2
\, \vert \eta(U)\vert^4\Big]
-\frac{{\overline \beta}_i}{4 \pi} 
\ln\bigg \vert \frac{\vartheta_1(\Delta_{\rho_2}-U
 \Delta_{\rho_1}\vert U)}{\eta(U)^3}\bigg\vert^2\nonumber\\
\nonumber\\
&+&\frac{{\overline \beta}_i}{4 \pi} \Big[
\ln\vert 2\pi (\rho_2-U \rho_1)\vert^2
+2\pi \vert U_2\vert \Delta_{\rho_{1}}^2\Big],
\end{eqnarray}
valid if
\begin{equation}\label{vevs}
\max \left\{\frac{1}{R_1}, \frac{1}{R_2 \sin\theta}, 
\frac{\vert\rho_1\vert}{R_1},
\frac{\vert\rho_2 - U \rho_1\vert}{R_2 \sin\theta}\right\}
\ll \Lambda
\end{equation}
The first two conditions in (\ref{vevs}) follow from (\ref{bound_xi}). The 
last two conditions originate from imposing that  the argument of
the $\Gamma$ functions present in  eqs.(\ref{j03}), (\ref{cL123}) be 
small enough, to approximate these functions with the familiar 
logarithm\footnote{We use $-\Gamma[0,z]=\gamma+\ln z+\cdots,\, 0<z\ll
1$, and $\gamma=0.577216...$ is the Euler constant.} of one-loop 
radiative corrections (these conditions can thus be relaxed). 
They state that the vev's of the Wilson lines in any 
directions in the root space be smaller than
$\Lambda$:  $v_\sigma^2, w_\sigma^2\ll\Lambda^2$, (with 
definition (\ref{defvw})).

The first term in (\ref{omega_t}) is a correction due to $(0,0)$ modes. The 
second term is due to non-zero modes and shows that  the leading UV behaviour 
(i.e. $T_2^*$ dependence) of the correction $\Omega_i$ is  not changed 
from the case of vanishing  vev of the Wilson line.  Indeed, this term  has
a  power-like dependence  $\Omega_i\sim T^*_2=\Lambda^2 R_1 R_2 \sin\theta$
and a logarithmic one  $\ln T_2^*\propto \ln \Lambda$ 
similar  to those of Case~1. Since the Wilson lines  introduce 
a  spontaneous  gauge  symmetry breaking, this confirms the 
expectation that the  UV behaviour of the couplings not be 
changed by their acquiring of non-zero vev's. The last  two  terms  in 
(\ref{omega_t}) are due to the Wilson lines effects alone. 
These terms depend on $U$  and  $\rho_{1,2}$ or rather its fractional part 
$\Delta_{\rho_{1,2}}$,  but no additional UV scale dependence is
present.

There is one notable difference between $\Omega_i$ of (\ref{omega_t})
and  $\Omega_i$ of Case 1 or Case 2 (A). This is  the absence 
in  (\ref{omega_t}) of the term
\begin{equation}\label{discontinuity}
(Q^2 R_1 R_2 \sin\theta) \ln \Big[ 4 \pi e^{-\gamma} 
{U_2}/{T_2^*}\Big]
\end{equation}
This term was present in Cases 1, 2 (A)  due to  massless
Kaluza-Klein states and it was induced by the contribution 
of massive Kaluza-Klein states  in the presence of the infrared
regulator required by the  massless modes. Such term 
is not present in (\ref{omega_t}) since no massless states exist
in this case.

The result of eq.(\ref{omega_t})   can  be re-written in a  more 
compact form
\begin{equation}\label{equiv_form}
\Omega_i\!=\frac{\beta_i}{4\pi} \ln \frac{\Lambda^2 (R_2
\sin\theta)^2}{\pi e^\gamma \vert \rho_2 -U \rho_1\vert^2}
+
\frac{\overline \beta_i}{4\pi} \bigg[T_2^* 
-
\ln\frac{T_2^* U_2 \, e^{-\gamma}}{\pi \vert \rho_2-U \rho_1\vert^2}
-
\ln\bigg\vert \frac{\vartheta_1(\Delta_{\rho_2}-U\Delta_{\rho_1}\vert U)}
{\eta(U)}\bigg\vert^2\!\!
\! +
2\pi \, U_2 \Delta_{\rho_1}^2\bigg]
\end{equation}
This is  the main result of the paper and gives the general 
correction to the gauge couplings in the presence of background gauge 
fields of vev's  which break the gauge symmetry to $\cG^*$.

We note that the results (\ref{omega_t}), (\ref{equiv_form}) 
were evaluated  for $\rho_1$ non-integer, while $\rho_2$ was kept  arbitrary. 
Further, taking in (\ref{omega_t}) the  formal limit $\rho_1$ integer 
($\Delta_{\rho_1}=0$) with $\rho_2$ non-integer gives a finite result.
Detailed calculations in Appendix  \ref{case3} show that this limit
does provide the correct result. In conclusion one can extend  the validity of
eqs.(\ref{omega_t}), (\ref{equiv_form}) to all cases with  either  
$\rho_1$ or $\rho_2$  non-integer and real values of the other, 
respectively. This is partly expected given the somewhat
symmetric role that $\rho_1$ and $\rho_2$ play.

In the formal limit when  both $\rho_1$ and $\rho_2$ are
integers (for a given $\sigma$)  the radiative correction 
$\Omega_i$  of (\ref{omega_t}), (\ref{equiv_form}) becomes  
logarithmically divergent in the  IR region and a regulator 
is required. In such situations  to evaluate $\Omega_i$ 
one should  apply the approach of Case 2 (A).  
This shows that  $\Omega_i$ as a function of $\rho_{1,2}$ is not
continuous at points in the ``moduli'' space of $(\rho_1,\rho_2)$ with 
($\rho_1,\rho_2)=(m,n)$ with $m,n \in \bZ$ when massless Kaluza-Klein
states may appear. Recalling that $\rho_i$ are $\sigma$ dependent, 
such states would  signal for $\sigma=\alpha$ an enhancement of the gauge 
symmetry beyond $\cG^*$. 
This discontinuity is also due to the term in eq.(\ref{discontinuity})
present for $\rho_{1,2}\in \bZ$.
To conclude, values of the Wilson lines vev's corresponding to 
$\rho_{i}$ integers cannot be smoothly reached {\it perturbatively} 
from those with $\rho_{i}$ non-integers.

We can now address the link of the result (\ref{equiv_form}) for
the one-loop corrections in the presence of Wilson lines 
with (heterotic) string theory. When doing so we only refer to the
term in (\ref{equiv_form}) proportional to $\overline\beta_i$ and 
due to {\it non-zero} modes, and which is evaluated at the string level.
The result (\ref{equiv_form}) has a  string counterpart 
in the context of  heterotic   (0,2) string compactifications 
\cite{Mayr:1995rx} with N=2 sector and Wilson line background which 
breaks the initial gauge  symmetry.  To compare the two results one
must consider the string result in the limit of large compactification 
area in string units,   or equivalently  the limit $\alpha'\!\ra\! 0$.  
The string correction to the gauge coupling  due to {\it non-zero} 
levels of Kaluza-Klein and winding modes  is in this limit
\cite{Mayr:1995rx}
\begin{equation}\label{hstring}
\Omega_{i,H}=\frac{\overline\beta_{i,H}}{4\pi}\left[
\frac{2\pi}{5} T_{2} - \ln(T_2 U_2) - \ln\vert\eta(U)\vert^4 - 
\frac{1}{5}\ln 
\bigg\vert\frac{\vartheta_1(B \vert U)}{\eta(U)}\bigg\vert^2\right]_H
\end{equation}
where the index $H$ stresses that the notation used in this
equation is  that of the  heterotic string, to be distinguished from 
that used in our field theory approach.   $T_2$ of (\ref{hstring})
is the imaginary part of the K\"ahler structure measured in $\alpha'$  
units and  has a field theory  counterpart in $T_2^*$ measured in UV cut-off 
length unit and defined in (\ref{moduli}), (\ref{cutoff}). 
Also $B=\mu-U\mu'$ is a mass shift induced by
$\mu,\mu'$ which are the two Wilson lines vev's at the string level
in a particular direction in the root space of the group considered there. 
$\mu,\mu'$ have field theory counterparts in $\rho_{1,2}$ present in 
eqs.(\ref{symmetry}), (\ref{wilsonvev}), (\ref{kkmass}).

Comparing  (\ref{equiv_form}), (\ref{hstring}) one notices that the
power-like and logarithmic behaviour in $T_2^*$ and  $T_2$
respectively is similar, up to the numerical coefficient which 
in the effective field theory approach is regularisation  
dependent ($T_2^*$ depends on UV regulator  $\xi$)  and cannot be 
fixed on field theory grounds only \cite{Ghilencea:2002ff}. 
The presence of the odd elliptic theta 
function $\vartheta_1$ is similar in (\ref{equiv_form}) and 
(\ref{hstring}). However,  its argument  depends 
at the field theory level on $\Delta_{\rho_{1,2}}$ which is the 
Wilson lines vev's $\rho_{1,2}$ modulo an
integer, as actually expected from eq.(\ref{symmetry}). 
Further, the term $\ln\vert\eta(U)\vert^4$  appearing  in (\ref{hstring})
and not present in (\ref{equiv_form}) is not a discrepancy of
the two approaches. This term is due at the string level to 
including the contribution $\mu=\mu'=0$. At the  field theory 
level this  requires one consider the case $\rho_{i,\alpha}=0$,
not included in (\ref{equiv_form}) and  which does bring in such a 
term as shown in  Case~1  eqs.(\ref{finalresult}).
The only difference between the two approaches 
is the existence  in (\ref{equiv_form})  of the term
$\ln\vert\rho_2-U\rho_1\vert$, absent in (\ref{hstring}).
With this exception,  the effective  field theory and  string
calculation in the 
limit $\alpha'\!\ra \!0$  lead to remarkably close results, despite
their entirely different approaches to computing $\Omega_i$. 

We conclude with a reminder  that the most general  correction to the
gauge couplings in the presence of  Wilson lines is a 
sum of the results of type found  in 
Cases 1 and 2 (B) (ignoring the very 
special case of integer Wilson line vev's $\rho_{k,\alpha}$). 
 According to the discussion
in Section  \ref{wilson_c} and 
eq.(\ref{QFTthresholds1}) the total correction  is a sum 
of eq.(\ref{omega_t}) and  eq.(\ref{finalresult}). 
Eq.(\ref{omega_t}) is associated with ``broken'' generators of initial $\cG$
while  eq.(\ref{finalresult}) is 
associated with ``unbroken'' generators $E_\alpha$ of $\cG^*$ (with
appropriate beta functions). Note 
 that $\beta_i$, $\rho_{i}$, $\Delta_{\rho_i}$
have all a  dependence  on $\sigma=\alpha,\lambda$ not shown explicitly
in this section.

\section{Conclusions.}\label{conclusions}

In general models with additional compact dimensions have 
a larger amount of gauge symmetry  than the SM  does
and a mechanism to break it is required. 
A natural procedure to achieve this is that of
Wilson line breaking   in which  components 
of the higher  dimensional gauge  fields develop vacuum expectation 
values in some directions in the root space. 
It is interesting to note that the effect of the background 
(gauge) field may be re-expressed as a ``twist'' of the  boundary conditions 
of the initial fields (with respect to the compact dimensions)
and which is removed by the formal limit of vanishing Wilson lines vev's.
A consequence of this symmetry breaking  is that after  
compactification (part of) the  4D Kaluza-Klein mass spectrum of the 
initial  fields is changed and the   levels 
are shifted by values proportional to the compactification radii and
the Wilson lines vev's.

We considered  generic 4D N=1 models with one and  two extra 
dimensions compactified on a circle and two-torus 
respectively,  in the presence of a constant  background
(gauge) field. 
Such models are ``field theory'' ($\alpha'\!\ra\! 0$)  limits of  
4D N=1 orbifold 
compactification of the  heterotic string with an  N=2 sector 
(``bulk'') in the  presence of Wilson lines.
For these  models we evaluated the  structure of the overall 
one-loop correction to the 4D gauge couplings including Wilson line effects.

The results depend significantly  on the directions (in the root space) 
of the vev's of the Wilson lines which in a realistic  model are
expected to be fixed by a dynamical (possibly non-perturbative)
mechanism. We computed the corrections to the gauge
couplings for cases when the initial  gauge symmetry is broken to 
a subgroup $\cG^*$  and for the special  case when 
a non-zero level Kaluza-Klein state of the tower  becomes massless,
leading to an enhancement of the gauge symmetry.

The calculation required a careful  analysis of the  UV and IR
behaviour  of the gauge couplings. 
The Wilson line corrections were identified and it was observed 
that the UV behaviour of the models considered is not worsened 
by their  non-zero vev's. The couplings (regarded as functions of 
the Wilson line vev's $\rho_{k,\alpha}$) 
have a  discontinuity  (infrared divergence) at all ``moduli'' 
points where   Kaluza-Klein states of non-zero  levels become massless. 
As a result, values of the Wilson lines  vev's corresponding to 
$(\rho_{1,\sigma},\rho_{2,\sigma})$ 
integers ($\sigma$ fixed)  cannot be smoothly reached {\it perturbatively} 
from those with $(\rho_{1,\sigma}, \rho_{2,\sigma})$ non-integers.

The results obtained were compared with their heterotic  
string counterpart. When no massless state is present in a
Kaluza-Klein tower (Case 2 (B)), the one-loop correction of the
effective field theory approach has 
strong similarities with that of the  heterotic string  
in the presence of Wilson lines and  in the 
 limit $\alpha'\ra 0$ (when the effects of 
winding modes are negligible).  This finding is remarkable given
the different approach of the two methods and shows that effective
field theories  can indeed yield very reliable results in the region
of large compactification radii (in units of UV cut-off length).
In both cases the results can be written as a sum over elliptic 
theta  functions of genus one (at the string level the general 
correction is further  associated
with the topology of a  genus two Riemann surface).
When a massless state is present in a Kaluza-Klein tower
an infrared regulator is needed for evaluating the contribution
to the gauge couplings (Cases 1, 2 (A)). For two compact dimensions
  this has as effect the 
presence of a correction which cannot be
recovered by the (infrared regularised) string calculations available, 
in the limit     $\alpha'\!\!\ra\!  0$. This is ultimately
caused by the infrared regularisation of the string which does not commute
with the (field theory) limit $\alpha'\ra 0$.

The techniques developed in this work can easily be applied to
specific models. The results obtained may be used for  the 
study of the  unification of the gauge couplings 
in MSSM-like models derived  from a grand unified model with Wilson 
line gauge symmetry breaking. The results can also be applied 
to  models with extra dimensions with  a  structure of the
4D Kaluza-Klein masses similar to the general 
one considered in this paper, even when this structure is not induced by a
Wilson line background, but by orbifolding, etc.   
It would  be interesting to know 
if the results of this work can be extended to regions of small 
radii of compactification (in units of UV cut-off length) and if 
the agreement  with the corresponding string results can be 
maintained.

\section*{\bf Acknowledgements:}
The author thanks Graham Ross and Fernando Quevedo for many
discussions on related topics. This  work was supported by a
post-doctoral research fellowship from PPARC, United Kingdom. 
Part of this work was performed during a 
visit at  CERN   supported by the RTN-European Program 
HPRN-CT-2000-00148, {\it ``Physics Across the Present Energy Frontier: 
Probing the Origin of Mass''.}

\newpage
\section{Appendix}
\appendix
\def\theequation{\thesection-\arabic{equation}} 
\section{Evaluation of Kaluza-Klein Integrals.}
 \setcounter{equation}{0}
\label{computing_1}

In the following a ``primed'' sum $\sum_{m}' f(m)$ stands for a sum 
over all non-zero,  integer numbers, $m\in\bZ-\{0\}$.
Similarly, $\sum_{m,n}' f(m,n)$ is a sum over all integers 
$(m,n)$ with $(m,n)\not=(0,0)$.

%%%%%%%%%%%%%%%%%%%%%%%%%%%%%%%%%%%%%%%%%%%%%%%%%

\subsection{Computing $\cR_1$}\label{r1ap}
$\bullet$ We compute the integral ($\xi>0$, $\delta>0$)
\begin{equation}\label{R1}
\cR_1 [\xi,\rho,\delta] 
\equiv \int_{\xi}^{\infty} \frac{dt}{t} \sum_{m}' e^{-\pi t
[ (m+\rho)^2 + \delta]}
\end{equation}
for $\xi\ll 1$, $\rho\in \bR$. With this notation, 
the  integral encountered  in the text in eq.(\ref{omegai1}) 
will be given by $\cJ=\cR_1[\xi/(R\mu)^2,
\rho,\chi (R \mu)^2]$
while $\cL_1$ in eq.(\ref{cL123}) is  
$\cL_1=\cR_1[\xi/(T_2 U_2),\, \rho_2-U_1 \rho_1,\, U_2^2 \rho_1^2]$.

To compute $\cR_1$ we use the Poisson re-summation formula
\begin{equation}\label{p_resumation}
\sum_{n\in Z} e^{-\pi A (n+\sigma)^2}=\frac{1}{\sqrt A} \sum_{\tilde
n\in Z} e^{-\pi A^{-1} \tilde n^2+2 i \pi \tilde n \sigma}
\end{equation}
We have
\begin{eqnarray}
\!\!\!\!\!\!\!\!\!\!\!\!\cR_1 [\xi,\rho,\delta] &=&
\int_{\xi}^{\infty}\frac{dt}{t} 
\bigg[- e^{-\pi t \rho^2}+\sum_{m} e^{-\pi \, t (m+\rho)^2} \bigg]
e^{-\pi\delta\,t} \\
&=&
\int_{\xi}^{\infty}\frac{dt}{t} \bigg[-e^{-\pi t
\rho^2}+\frac{1}{\sqrt{t}}+
\frac{1}{\sqrt{ t}}\sum_{m}'e^{-\pi \tilde m^2/t+2 i\pi \tilde m
\rho}\bigg] \, e^{-\pi\delta \, t}\label{R1_1}\\
&=&\!\!
\frac{2 e^{-\pi \xi \delta} }{\sqrt { \xi}} 
+2 \pi \sqrt{\delta } \textrm{Erf}[\sqrt{\pi \xi \delta}]
-\Gamma[0,\pi\xi(\rho^2+\delta )]
-\ln \Big\vert 2 \sin \pi (\rho+i\sqrt\delta )\Big\vert^2\!, \,\,\,
\xi\ll1 
\label{R1_2}
\end{eqnarray}
For the last term in (\ref{R1_1})  we assumed $\xi\ll 1$ and we 
used that \cite{gr}
\begin{equation}\label{bessel1}
\!\!\! \!\!\!
\int_{0}^{\infty} \! dx\, x^{\nu-1} e^{- b x^p- a
x^{-p}}=\frac{2}{p}\, \bigg[\frac{a}{b}
\bigg]^{\frac{\nu}{2 p}} K_{\frac{\nu}{p}}(2 \sqrt{a \, b}),\quad Re
(b), Re (a)>0; \quad
K_{-\frac{1}{2}}(z)=\sqrt{\frac{\pi}{2 z}} e^{-z}
\end{equation}
In the last integral in (\ref{R1_1}) we set $\xi=0$ (the integral
has no divergence in $\xi$ or in $\delta\ra 0$ e.g. 
$\xi\ln \delta$). Indeed the error  $\epsilon_1$ induced by doing so vanishes
\begin{eqnarray}
\vert\epsilon_1\vert&=&\bigg\vert
\int_{0}^{\xi}\frac{dt}{t^{3/2}} \sum_{\tilde m}' e^{-\pi \tilde
m^2/t+2 i \pi \tilde m \rho} \,e^{-\pi \delta t} \bigg\vert
\leq 
\sum_{\tilde m}' \int_{1/\xi}^{\infty} \frac{dt}{t^{1/2}} 
e^{-\pi \tilde m^2 t} 
\leq
2 \sum_{\tilde m>0} \int_{1/\xi}^{\infty} \frac{dt}{t^{1/2}} 
e^{-\pi \tilde m t}\nonumber\\ 
&\leq & 
\frac{2}{1- e^{-\pi/\xi}} \int_{1/\xi}^{\infty}\frac{dt}{t^{1/2}}
e^{-\pi t} 
\leq 
\frac{2 \xi}{1- e^{-\pi/\xi}} \bigg[\frac{3}{2\pi e}\bigg]^{\frac{3}{2}}
\ll 1
\quad \textrm{if} 
\quad \xi\ll 1\label{delta1}
\end{eqnarray}
Thus $\vert\epsilon_1\vert$ 
vanishes if $\xi\!\ll\! 1$ for $\delta>0$, $\rho\in\bR$
and the result for
$\cR_1$ is indeed  that given by eq.(\ref{R1_2}).

\noindent
Further, with the expansions  \cite{gr}
\begin{eqnarray}\label{definition}
\textrm{Erf}[x]& \equiv& \frac{2}{\sqrt\pi}\int_{0}^x dt \,
e^{-t^2}=\frac{2 x}{\sqrt \pi}-\frac{2 x^3}{3\sqrt \pi}+\cO(x^5),\qquad
\textrm{if  } x\ll 1,\nonumber\\
-\Gamma[0,z]&=& \gamma+\ln z+\sum_{k\geq 1}\frac{(-z)^k}{k! \, k},
\qquad \textrm{for  }  z>0
\end{eqnarray}
we obtain an approximation for $\cR_1$ of  eq.(\ref{R1_2}) in
the limit $\xi\ra 0$
\begin{eqnarray}\label{a10}
\cR_1[\xi,\rho,\delta] =  -\ln\bigg[4 \pi e^{-\gamma}
\frac{1}{\xi}\, e^{-2/\sqrt{\xi}}\bigg]
-
\ln\bigg\vert  \frac{\sin \pi(\rho+i \sqrt \delta)}
{\pi(\rho+ i \sqrt \delta)}\bigg\vert^2,\quad 
\xi\ll \min\left\{1, \frac{1}{\pi \delta}, \frac{1}{\pi
(\rho^2+\delta)}\right\}
\end{eqnarray}
If $\rho$ is non-integer 
($\rho\not \in \bZ$)  and  $\delta=0$, one finds from (\ref{R1_2})
\begin{equation}\label{limit1}
\cR_1[\xi,\rho\not\in \bZ,0]=\int_{\xi}^{\infty}\frac{dt}{t}
\sum_{m}' e^{-\pi \, t (m+\rho)^2}
=\frac{2}{\sqrt{\xi}}-\Gamma[0,\pi\xi\rho^2]-\ln \vert 2 \sin\pi
\rho\vert^2,\qquad \xi\ll 1
\end{equation}
If $\rho$ is a non-zero integer ($\rho\in \bZ^*$) 
\begin{equation}\label{a12}
\cR_1[\xi,\rho\in\bZ^*,\delta]=\frac{2}{\sqrt\xi}-
\Gamma[0,\pi\xi\rho^2]-\ln\vert2\sin\pi i
\sqrt\delta\vert^2, \qquad \xi\ll 1,\, \delta\ll 1
\end{equation}
If $\rho\ra 0$ and $\delta\ra 0$ (the limits commute) 
\begin{eqnarray}\label{limit2}
\cR_1[\xi,0,0]=\int_{\xi}^{\infty}\frac{dt}{t}
\sum_{m}' e^{-\pi \, t m^2}
=-\ln \Big[ 4\pi e^{-\gamma} \frac{1}{\xi} e^{-2/\sqrt{\xi}}\Big],\quad
\qquad \xi\ll 1
\end{eqnarray}
Eq.(\ref{R1_2}) and (\ref{a10}) were used in the text,
eq.(\ref{omegai1}), (\ref{cL123}).

\noindent
For related results on Kaluza-Klein integrals see also the 
Appendix in \cite{Ghilencea:2002ff} and \cite{Ghilencea:2002ak}.

%%%%%%%%%%%%%%%%%%%%%%%%%%%%%%%%%%%%%%%%%%%%%%%%%%%%%%%%%%%%%%

\vspace{0.55cm}
\subsection{Computing $\cR_2$}\label{r2ap}
$\bullet$ We compute the integral:
\begin{equation}\label{R2}
\cR_2 [\xi,\rho] \equiv \int_{\xi}^{\infty} \frac{dt}{t^{3/2}} \sum_{m}'
e^{-\pi t (m+\rho)^2}
\end{equation}
for $\xi\ra 0$ and $\rho\not \in \bZ$. 
With this notation, integral $\cL_2$ in eq.(\ref{cL123}) is  given by 
$\cL_2=\cR_2[\xi U_2/T_2, \rho_1] \, \vert U_2\vert$.

The constant $\rho$ can be written as
\begin{equation}
\rho=[\rho]+\Delta_\rho,\quad \textrm{with}\quad
[\rho]\in \bZ, \qquad 0<\Delta_\rho <1
\end{equation}
$\Delta_\rho$ is the fractional part of $\rho$, positive definite, 
irrespective of the sign of  $\rho$. $\rho\not\in \bZ$ thus
$\Delta_\rho\not=0$.

\noindent
With (\ref{p_resumation}) one has
\begin{eqnarray}
\!\!\!\!\!\!\!\!\!\!\!\!\!\!
\cR_2[\xi,\rho]
& =&
\int_{\xi}^{1} \frac{dt}{t^{3/2}} \Bigg[
\frac{1}{\sqrt t}\sum_{\tilde m}' e^{-\pi \tilde m^2/t+2 i \pi \tilde m\rho}
+\frac{1}{\sqrt t}-e^{-\pi t \rho^2}\bigg]
 +  
\int_{1}^{\infty} \frac{dt}{t^{3/2}} \sum_{m}'
e^{-\pi t (m+\rho)^2}\nonumber\\
&=&
\int_{\xi}^{1}\frac{dt}{t^{3/2}} 
\bigg[\frac{1}{\sqrt t} - e^{-\pi t \rho^2} \bigg] 
+\int_{1}^{1/\xi} dt \sum_{\tilde m}' e^{-\pi\,t\,\tilde m^2+
2 i \pi \tilde m \rho}
 +  
\int_{1}^{\infty} \frac{dt}{t^{3/2}} \sum_{m}'
e^{-\pi t (m+\rho)^2}\label{error2}\nonumber\\
&=& \cI_\xi
+\int_{1}^{\infty} dt \sum_{\tilde m}' e^{-\pi\,t\,\tilde m^2+
2 i \pi \tilde m \rho}
 +  
\int_{1}^{\infty} \frac{dt}{t^{3/2}} \sum_{m}'
e^{-\pi t (m+\rho)^2}
\equiv \cI_\xi+\cF,\quad \xi\ll 1
\label{ixi}
\end{eqnarray}
We  introduced $\cI_\xi$
\begin{eqnarray}
\cI_\xi \equiv \int_{\xi}^{1}\frac{dt}{t^{3/2}} 
 \bigg[\frac{1}{\sqrt t} - e^{-\pi t \rho^2} \bigg] 
=\frac{1}{\xi}-\frac{2}{\sqrt \xi} e^{-\pi \rho^2 \xi}+2 e^{-\pi \rho^2}-1
+2\pi \rho \Big[\textrm{Erf}[\rho\sqrt \pi]-
\textrm{Erf}[\rho\sqrt{\pi \xi}]\Big]\label{ixii}
\end{eqnarray}
and  {\it finite} $\cF$ (since $\rho$ is non-integer)
\begin{eqnarray}
\cF & \equiv &  
\int_{1}^{\infty} dt 
\sum_{\tilde m}' e^{-\pi\,t\,\tilde m^2+ 2 i \pi \tilde m \rho}
 +  
\int_{1}^{\infty} \frac{dt}{t^{3/2}} \sum_{m}' e^{-\pi t (m+\rho)^2}
\nonumber\\
&=&\lim_{\epsilon \ra 0}
\int_{1}^{\infty}  t^\epsilon dt 
\sum_{\tilde m}' e^{-\pi\,t\,\tilde m^2+ 2 i \pi \tilde m \rho}
 +  
\int_{1}^{\infty} \frac{dt}{t^{3/2+\epsilon}} \sum_{m}' e^{-\pi t (m+\rho)^2}
\nonumber\\
&=&\lim_{\epsilon \ra 0}
\int_{0}^{1} \frac{dt}{t^{2+\epsilon}}\sum_{\tilde m}' 
e^{-\pi \tilde m^2/t +2 i \pi \tilde m \rho}
 +  
\int_{1}^{\infty} \frac{dt}{t^{3/2+\epsilon}} \sum_{m}' e^{-\pi t (m+\rho)^2}
\nonumber\\
&=&\lim_{\epsilon \ra 0}
\int_{0}^{\infty} \frac{dt}{t^{3/2+\epsilon}}\sum_{m}' 
e^{-\pi t (m+\rho)^2} 
+ \int_{0}^{1}\frac{dt}{t^{2+\epsilon}}
\Big[-1+\sqrt t \, e^{-\pi t \rho^2}\Big] \equiv \cG+\cH
\label{qqq}
\end{eqnarray}
where
\begin{eqnarray}
\cG &\equiv & \lim_{\epsilon \ra 0}
\int_{0}^{\infty} \frac{dt}{t^{3/2+\epsilon}}\sum_{m}' 
e^{-\pi t (m+\rho)^2} 
=\lim_{\epsilon \ra 0}
\int_{0}^{\infty} \frac{dt}{t^{3/2+\epsilon}}
\Big[\sum_{m} e^{-\pi t(m+\rho)^2} -e^{-\pi t \rho^2}\Big]\label{first}
\\
&=&\lim_{\epsilon \ra 0}
\int_{0}^{\infty} \frac{dt}{t^{3/2+\epsilon}}
\Big[e^{-\pi t \Delta_\rho^2} - e^{-\pi t \rho^2}
+ \sum_{m>0} e^{-\pi t (m+\Delta_\rho)^2}+
 \sum_{m>0} e^{-\pi t (m-\Delta_\rho)^2}\Big]\label{second}
\\
&=&\lim_{\epsilon \ra 0}
\pi^{\frac{1}{2}+\epsilon}\Gamma[-1/2-\epsilon]
\Big[\Delta_\rho^{1+2 \epsilon} -
 \vert\rho\vert^{1+2\epsilon}+
 \sum_{m\geq 0} {(m+1+\Delta_\rho)^{1+2\epsilon}}
+(\Delta_\rho\ra -\Delta_\rho)\Big]\label{third}
\\
&=&\lim_{\epsilon \ra 0}
2\pi (\vert\rho\vert -\Delta_\rho)
+\pi^{\frac{1}{2}+\epsilon}\Gamma[-1/2-\epsilon]\Big[
\zeta[-1-2\epsilon,1+\Delta_\rho]+
\zeta[-1-2\epsilon,1-\Delta_\rho]\Big]\\
\vspace{0.3cm}
&=& 2\pi \big\vert\rho\big\vert 
+2\pi B_2[\Delta_\rho]
\label{last}
\end{eqnarray}
In  (\ref{second}), (\ref{third})
we used that $\Delta_\rho\not=0$ (since  
$0<\Delta_\rho<1$ ($\rho\not\in\bZ$)) and  
\begin{equation}
\zeta[z,q]=\sum_{n\geq 0}(q+n)^{-z}
,\qquad \zeta[-1,x]=-({1}/{2}) \, B_2[x], \qquad
B_2[x]\equiv {1}/{6}+x^2-x,
\end{equation}
Also
\begin{equation}
\cH\equiv \lim_{\epsilon\ra 0} \int_{0}^{1}\frac{dt}{t^{2+\epsilon}}
\Big[-1+\sqrt t \, e^{-\pi t \rho^2}\Big]
=1+ \vert\rho\vert \sqrt\pi \Big[
\Gamma[-1/2]-\Gamma[-1/2,\pi\rho^2]\Big]
\label{cH}
\end{equation}
Adding together eq.(\ref{last}), (\ref{cH}) 
and using eq.(\ref{qqq}) gives
\begin{eqnarray}\label{fl}
\cF= 2\pi \vert \rho \vert 
+2\pi B_2[\Delta_\rho]
+1+\vert\rho\vert \sqrt\pi \Big[\Gamma[-1/2]-\Gamma[-1/2,\pi \rho^2]\Big]
\end{eqnarray}
Eqs.(\ref{ixi}), (\ref{ixii}) and  (\ref{fl}) give
\begin{eqnarray}\label{R2_final}
\cR_2[\xi,\rho]&=& \frac{1}{\xi}-\frac{2}{\sqrt \xi} e^{-\pi \rho^2 \xi}
-2\pi \rho \,\textrm{Erf}[\rho\sqrt{\pi\xi}]+2\pi\Big[\vert\rho\vert+
\frac{1}{6}+ \Delta_{\rho}^2-\Delta_{\rho}\Big], \nonumber\\
\nonumber\\
 \textrm{with:}  &&\quad \xi\ll 1,\, \, 
\rho\not \in \bZ, \,\,  (0<\Delta_\rho < 1)
\end{eqnarray}
\noindent
$\bullet$ We evaluate the error introduced in eq.(\ref{error2}),
while computing $\cR_2$ to ensure it vanishes for $\xi\!\ll\! 1$:
\begin{equation}\label{delta2}
\vert\epsilon_2\vert \equiv \bigg\vert \int_{1/\xi}^{\infty}
dt \sum_{m}' e^{-\pi \, t m^2+2 i \pi m \rho}\bigg\vert
\leq 2 \int_{0}^{\xi} \frac{dt}{t^2} \sum_{m>0} e^{-\pi m/t}
=\frac{-2}{\pi}\ln\big[1-e^{-\pi/\xi}\big]
\ll 1,\,\,\, \textrm{if}\,\,\,  \xi\ll 1
\end{equation}
Eqs.(\ref{R2_final}), (\ref{delta2}) were used in the text,
eq.(\ref{cL123}) and (\ref{bound_xi}).

%%%%%%%%%%%%%%%%%%%%%%%%%%%%%%%%%%%%%%%%%%%%%%%%%%%%%%%%%%%%%%%%%%%%%%%%%

\vspace{0.4cm}
\subsection{Computing $\cR_3$}\label{r3ap}
$\bullet$ We compute the integral:
\begin{equation}\label{R3}
\cR_3[\xi,\rho_1,\rho_2] \equiv \int_{\xi}^{\infty}
\frac{dt}{t^{3/2}}\sum_{m_1}'\sum_{\tilde m_2}'
e^{-\pi \tilde m_2^2/t-\pi \, t \, U_2^2 (m_1+\rho_1)^2 +2 i \pi \tilde
m_2 [\rho_2 -U_1 (\rho_1+m_1)]}
\end{equation}
for $0<\xi\ll 1$,  $U_{1,2}\not=0$, $\rho_1\not \in \bZ$ and real. 
Therefore $\cL_3$ of eq.(\ref{cL123}) is
$\cL_3=\cR_3[\xi/(T_2 U_2),\rho_1,\rho_2]$.

\vspace{0.3cm}
First we  introduce $\rho_1=[\rho_1]+\Delta_{\rho_1}$, 
where $[\rho_1]$ is an integer and
$\Delta_{\rho_1}$ is its fractional part defined as
$0<\Delta_{\rho_1}<1$.
To evaluate the integral of  $\cR_3$ we  use 
 eq.(\ref{bessel1})  after having set
$\xi=0$ in its lower limit (this is allowed for the integral of $\cR_3$ is
finite under the assumptions for which we evaluate it). 
Setting $\xi\!=\!0$ introduces a (vanishing) error in $\cR_3$ to be evaluated
 shortly (eq.(\ref{ddd})). One has
\begin{eqnarray}
\cR_3[\xi,\rho_1,\rho_2]&=& \sum_{m_1}' \sum_{\tilde m_2}'
 \frac{1}{\vert \tilde m_2\vert}
e^{2 i \pi \tilde m_2 (\rho_2- U_1 (\rho_1+m_1))}\, e^{-2\pi \vert \tilde
 m_2 (m_1+\rho_1) U_2\vert } \nonumber\\
&=&
\sum_{m_1} \sum_{\tilde m_2}'
 \frac{1}{\vert \tilde m_2\vert}
e^{2 i \pi \tilde m_2 (\rho_2- U_1 (\rho_1+m_1))} \, e^{-2\pi \vert \tilde
 m_2 (m_1+\rho_1) U_2\vert }\nonumber\\
&-&
\sum_{\tilde m_2}' \frac{1}{\vert\tilde m_2\vert}
e^{2 i \pi \tilde  m_2 (\rho_2 - U_1 \rho_1) -2\pi \vert 
\tilde m_2 \rho_1 U_2\vert}\nonumber\\
&=&
\sum_{m}' \sum_{\tilde m_2}' \frac{1}{\vert\tilde m_2\vert}
e^{2 i \pi \tilde m_2 (\rho_2 - U_1 (m+\Delta_{\rho_1}))-2 \pi 
\vert \tilde m_2 ( m+\Delta_{\rho_1}) U_2\vert}\nonumber\\
& + &
\bigg[\sum_{\tilde m_2}' \frac{1}{\vert \tilde m_2\vert} 
e^{2 i \pi \tilde m_2 (\rho_2 -U_1 \Delta_{\rho_1})
- 2 \pi \vert \tilde m_2 \Delta_{\rho_1}  U_2\vert}
-(\Delta_{\rho_1}\ra\rho_1)\bigg]
\end{eqnarray}
%\nonumber\\
where we used the notation $m=m_1+[\rho_1]\in \bZ$.

With $\vert m+\Delta_{\rho_1}\vert = \pm (m+\Delta_{\rho_1})$ for 
$m\geq 1$ and $m\leq -1$ respectively, one finds after some algebra
\begin{eqnarray}
\cR_3[\xi,\rho_1,\rho_2]
&=&
-\ln\prod_{m>0}\Big\vert 1-e^{2 i \pi m U} \, 
e^{-2 i \pi (\rho_2-\Delta_{\rho_1} U)}
\Big\vert^2 \, \Big\vert 1- e^{2 i \pi m
 U} \, e^{2 i \pi (\rho_2 -\Delta_{\rho_1} U)}
\Big\vert^2
\nonumber\\
&-&2\pi\vert U_2\vert
(\vert \rho_1\vert-\Delta_{\rho_1})
+\ln \bigg \vert \frac{\sin \pi(\rho_2 - U \rho_1)}
{\sin \pi (\rho_2 - U \Delta_{\rho_1})}\bigg\vert^2\label{R3_1}
\end{eqnarray} 
This result  may also be written as
\begin{eqnarray}
\!\!\cR_3[\xi,\rho_1,\rho_2]\!=\! -\ln \bigg \vert
\frac{\vartheta_1(\rho_2-\Delta_{\rho_1} U\, \vert\, U)}{\eta(U)}\bigg\vert^2
- 2 \pi \vert U_2\vert \Big[\frac{1}{6}-\Delta_{\rho_1}+\vert \rho_1\vert\Big]
+ \ln\Big \vert 2 \sin\pi (\rho_2
- U \rho_1)\Big\vert^2\label{R3_2}
\end{eqnarray}
with  the special functions $\eta$,  $\vartheta_1$
\begin{eqnarray}
\eta(\tau) & \equiv & e^{\pi i \tau/12} \prod_{n\geq 1} (1- e^{2 i
\pi\, n\, \tau}),\nonumber\\
\vartheta_1(z\vert\tau)&\equiv & 2 q^{1/8}\sin (\pi z) \prod_{n\geq 1} 
(1- q^n) (1-q^n e^{2 i \pi z}) (1- q^n e^{-2 i \pi z}), \qquad q\equiv e^{2
i \pi \tau}
%%  
%%  &=& \frac{1}{i}\sum_{n\in\bZ} (-1)^n e^{i\pi\tau (n+1/2)^2}\,
%%  e^{(2n+1)i\pi z} 
\label{R3_3}
\end{eqnarray}
%%  $\vartheta_1(z\vert\tau)$ is equal to $\vartheta_1(\pi z\vert\tau)$ of
%%  \cite{gr}, eq.8.180(2).

\noindent
$\bullet$
We now evaluate the error introduced by setting $\xi=0$ in the
integral for $\cR_3$. 
This error equals
\begin{eqnarray}
\!\!\!\!\!\!\!\!\!\!\!\!\!\!
\vert \epsilon_3\vert&\!\! \equiv\!\!  &\bigg\vert 
\int_{0}^{\xi}
\frac{dt}{t^{3/2}}\sum_{m_1}'\sum_{\tilde m_2}'
e^{-\pi \tilde m_2^2/t-\pi \, t \, U_2^2 (m_1+\rho_1)^2 +2 i \pi \tilde
m_2 [\rho_2 -U_1 (\rho_1+m_1)]}
\bigg\vert\nonumber \\
\!\!\!\!&\leq&\sum_{m_1}'\sum_{\tilde m_2}' 
\int_{0}^{\xi}
\frac{dt}{t^{3/2}}
e^{-\pi \tilde m_2^2/t -\pi t U_2^2 (m_1 +[\rho_1]+\Delta_{\rho_1})^2}
\nonumber\\
\!\!\!\!&=&
\sum_{m}'\sum_{\tilde m_2}'\int_{0}^{\xi}\frac{dt}{t^{3/2}} e^{-\pi
\tilde m_2^2/t-\pi t U_2^2 (m+\Delta_{\rho_1})^2}
+\bigg[\sum_{\tilde m_2}' \int_{0}^{\xi} \frac{dt}{t^{3/2}} e^{-\pi \tilde
m_2^2/t-\pi t\, U_2^2 \Delta_{\rho_1}^2}-(\Delta_{\rho_1}\!\!
\leftrightarrow\!\!{\rho_1})\bigg]
\label{ddd}
\end{eqnarray}
Each integral in the square bracket vanishes if $\xi\!\ll\! 1$. Indeed
with $\gamma$ standing for $\rho_1$ or $\Delta_{\rho_1}$ one has
\begin{eqnarray}
\cE_1
& \equiv &\sum_{\tilde m_2}' \int_{0}^{\xi} \frac{dt}{t^{3/2}} e^{-\pi \tilde
m_2^2/t-\pi t\, U_2^2 \gamma^2}
\leq 
2\sum_{\tilde m_2>0} \int_{0}^{\xi}\frac{dt}{t^{3/2}} e^{-\pi \tilde
m_2^2/t} 
\leq 
2 \sum_{\tilde m_2>0} \int_{0}^{\xi}\frac{dt}{t^{3/2}} e^{-\pi \tilde
 m_2/ t} \nonumber\\
&\leq & \frac{2}{1-e^{-\pi/\xi}}\int_{0}^{\xi}\frac{dt}{t^{3/2}}
e^{-\pi/t}
\leq \frac{2 \xi}{1-e^{-\pi/\xi}}\bigg[\frac{3}{2\pi
e}\bigg]^{3/2} \ll 1 \quad \textrm {if}\,\, \xi\ll 1.\label{error3}
\end{eqnarray}
Similarly, the first integral in (\ref{ddd}) is vanishing for $\xi$
small enough:
\begin{eqnarray}
\cE_2 &\equiv&
2 \sum_{m>0} \sum_{\tilde m_2>0} \int_{0}^{\xi} \frac{dt}{t^{3/2}}
e^{-\pi \tilde m_2^2/t}\bigg[e^{-\pi t U_2^2 (m+\Delta_{\rho_1})^2}
+(\Delta_{\rho_1}\leftrightarrow -
\Delta_{\rho_1})\bigg]\nonumber\\
&\leq & 2 \sum_{m>0} \sum_{\tilde m_2>0} \int_{0}^{\xi} \frac{dt}{t^{3/2}}
e^{-\pi \tilde m_2^2/t}\bigg[e^{-\pi t U_2^2 m^2}+
e^{-\pi t U_2^2 (m-1)^2}\bigg]\nonumber\\
&\leq & 4  \sum_{m>0} \sum_{\tilde m_2>0}
 \int_{0}^{\xi} \frac{dt}{t^{3/2}}
e^{-\pi \tilde m_2^2/t} e^{-\pi t U_2^2 m^2}
+2 \sum_{\tilde m_2>0}\int_{0}^{\xi} \frac{dt}{t^{3/2}} e^{-\pi \tilde
m_2^2/t}
\end{eqnarray}
The last integral was already shown to vanish for $\xi\ll 1$, while 
the first integral is smaller than 
\begin{eqnarray}
\!\!\!\!\!\!\!\!\!\!\!\!\!\!
4  \sum_{m>0} \sum_{\tilde m_2>0}
 \int_{0}^{\xi} \frac{dt}{t^{3/2}}
e^{-\pi \tilde m_2/t} e^{-\pi t U_2^2 m}
&=&4  \int_{0}^{\xi} \frac{dt}{t^{3/2}}
\frac{e^{-\pi/t}}{1-e^{-\pi/t}} \frac{e^{-\pi U_2^2 t}}{1-e^{-\pi
 U_2^2 t}}\nonumber\\
&\leq& 
\frac{4}{1-e^{-\pi/\xi}}\frac{\xi}{\pi U_2^2} \bigg[\frac{5}{2\pi e}
\bigg]^{5/2}\ll 1 \quad \textrm{ if} \,\, \xi\ll\textrm{min}\{1, U_2^2\}
\label{lastcond}
\end{eqnarray}
For the last factor under last integral 
we used that $e^{-a }/(1-e^{-a })<1/a $, $a>0$.
Eqs.(\ref{error3}), (\ref{lastcond}) set
the conditions for which the results for $\cR_3$, (\ref{R3_1}),
(\ref{R3_2}) hold true:
\begin{equation}
\vert \epsilon_3\vert \ll 1 \quad \textrm{ if } \,\,\, \xi\ll \min
\left\{1, U_2^2\right\}\label{delta3}
\end{equation}
Eqs.(\ref{R3_2}), (\ref{delta3}) were used  in the text,  
eqs.(\ref{cL123}), (\ref{bound_xi}).

%%%%%%%%%%%%%%%%%%%%%%%%%%%%%%%%%%%%%%%%%%%%%%%%%%%%%%%%%%%%%%%%%%%%%%%%%

\vspace{0.2cm}
\subsection{General Kaluza-Klein integrals (in DR and proper-time
    cut-off).}\label{r2general}

$\bullet$ A generic presence in models with compact dimensions is 
a generalised version of the integral  $\cR_2$
\begin{equation}\label{R2g}
\cR^* [\xi,\rho,\delta,\nu] 
\equiv \int_{\xi}^{\infty} \frac{dt}{t^{\nu}} \sum_{m}'
e^{-\pi t (m+\rho)^2 -\pi \delta t}, \qquad \xi,\,\delta,\, \nu>0
\end{equation}
with $\xi\!\ll 1$, $\delta\ll 1$, with $\nu>0$ and  $\rho$ real 
($\xi$ and $\delta$ are UV and IR regulators, respectively).
For future reference we outline the computation of $\cR^*$ 
and  of its DR version $\cG^*$, following the approach used for $\cR_2$ of
Appendix  \ref{r2ap}. 
One can have  $\delta=0$ provided that $\rho\not\in \bZ-\{0\}$.
We write $\rho$  as
\begin{equation}
\rho=[\rho]+\Delta_\rho,\quad \textrm{with}\quad
[\rho]\in \bZ, \qquad 0\leq \Delta_\rho <1
\end{equation}
with $\Delta_\rho$ the fractional part of $\rho$. Thus 
$\Delta_\rho^2+\delta\not=0$, unless $\rho=0=\delta\,(=\!\Delta_\rho)$.
With eq.(\ref{p_resumation}) one has
\begin{eqnarray}
\cR^*& =&
\int_{\xi}^{1} \frac{dt}{t^{\nu}} \bigg[\frac{1}{\sqrt t}\sum_{\tilde m}' 
e^{-\pi \tilde m^2/t+2 i \pi \tilde m\rho}
+\frac{1}{\sqrt t}-e^{-\pi t \rho^2}\bigg] e^{-\pi \delta t}
 +  
\int_{1}^{\infty} \frac{dt}{t^{\nu}} \sum_{m}'
e^{-\pi t (m+\rho)^2 -\pi \delta t }\nonumber\\
&=&
\int_{\xi}^{1}\frac{dt}{t^{\nu}} 
\bigg[\frac{1}{\sqrt t} - e^{-\pi t \rho^2} \bigg] e^{-\pi \delta t}
+
\int_{1}^{\frac{1}{\xi}} \! \frac{dt}{t^{3/2-\nu}}
\sum_{\tilde m}' e^{-\pi\,t\,\tilde m^2+2 i \pi \tilde m \rho-\pi \delta/t}
 +  
\int_{1}^{\infty} \frac{dt}{t^{\nu}} 
\sum_{m}'e^{-\pi t (m+\rho)^2-\pi \delta t}
\label{error2g}\nonumber\\
&=& \cI_\xi^*
+
\int_{1}^{\infty} \frac{dt}{t^{3/2-\nu}}
\sum_{\tilde m}' e^{-\pi\,t\,\tilde m^2+2 i \pi \tilde m \rho -\pi \delta/t}
 +  
\int_{1}^{\infty} \frac{dt}{t^{\nu}} 
\sum_{m}' e^{-\pi t (m+\rho)^2-\pi t\delta},\qquad \xi\ll 1
\nonumber\\
&\equiv &\cI_\xi^*+\cF^*,\qquad \xi\ll 1
\label{ixig}
\end{eqnarray}
with an obvious notation in the last steps.
 $\cI_\xi^*$ and $\cF^*$ are computed  below, eqs.(\ref{ixiig}) 
and (\ref{qqqg}). $\cF^*$ is {\it finite} within our assumptions on
$\delta$, $\rho$ ($\Delta_\rho$). Its integrand is always
exponentially suppressed. In the second line above, the integral  
on the interval  $(1,1/\xi)$ was actually evaluated on  $(1,\infty)$. 
This introduces an  error $\int_{1/\xi}^\infty \textrm{(integrand)}$ 
which can be shown to vanish as in eq.(\ref{delta1}) 
if  $\xi\!\ll\! 1$ and $\nu\!>\!0$.

\noindent
One has
\begin{eqnarray}
\cI_\xi^* \equiv \int_{\xi}^{1}\frac{dt}{t^{\nu}} 
 \bigg[\frac{1}{\sqrt t} - e^{-\pi t \rho^2} \bigg] e^{-\pi \delta t}
&=&
[\pi (\delta+\rho^2)]^{\nu-1}
\Big[\Gamma[1-\nu,\pi(\delta+\rho^2)]
-\Gamma[1-\nu,\pi(\delta+\rho^2)\xi]\Big]
\nonumber\\
&-&
(\pi\delta)^{\nu-1/2}
\Big[\Gamma[1/2-\nu,\pi\delta]-\Gamma[1/2-\nu,\pi\delta\xi]\Big]
\label{ixiig}
\end{eqnarray}
Further, with $\epsilon\!\ra\! 0$ and 
using (\ref{p_resumation}) one can re-write the  {\it finite} $\cF^*$ as
\begin{eqnarray}
\!\!\!\!\!\!\!\!
\cF^* & \equiv &  
\int_{1}^{\infty} \frac{dt}{t^{3/2-\nu}} 
\sum_{\tilde m}' e^{-\pi\,t\,\tilde m^2+ 2 i \pi \tilde m \rho-\pi\delta/t}
 +  
\int_{1}^{\infty} \frac{dt}{t^{\nu}} \sum_{m}' 
e^{-\pi t (m+\rho)^2-\pi \delta t}
\nonumber\\
&=&\lim_{\epsilon \ra 0}
\int_{1}^{\infty}  t^\epsilon \frac{dt}{t^{3/2-\nu}} 
\sum_{\tilde m}' e^{-\pi\,t\,\tilde m^2+ 2 i \pi \tilde m \rho-\pi\delta/t}
 +  
\int_{1}^{\infty} \frac{dt}{t^{\nu+\epsilon}} \sum_{m}' e^{-\pi t
(m+\rho)^2-\pi \delta t}
\nonumber\\
&=&\lim_{\epsilon \ra 0}
\int_{0}^{1} \frac{dt}{t^{1/2+\nu+\epsilon}}\sum_{\tilde m}' 
e^{-\pi \tilde m^2/t +2 i \pi \tilde m \rho-\pi \delta t}
 +  
\int_{1}^{\infty} \frac{dt}{t^{\nu+\epsilon}} \sum_{m}' e^{-\pi t
(m+\rho)^2-\pi \delta t}
\nonumber\\
&=&\lim_{\epsilon \ra 0}
\int_{0}^{\infty} \frac{dt}{t^{\nu+\epsilon}}
\sum_{m}' e^{-\pi t (m+\rho)^2-\pi \delta t} 
+ 
\int_{0}^{1}\frac{dt}{t^{1/2+\nu+\epsilon}}
\Big[\sqrt t \, e^{-\pi t \rho^2}-1\Big] e^{-\pi \delta t}
\equiv \cG^*\!+\!\cH^*
\label{qqqg}
\end{eqnarray}
The first integral (denoted $\cG^*$)  is  just a DR
version of $\cR^*$ and is evaluated below. 
One has ($\epsilon\!\ra \!0$)
\begin{eqnarray} 
\cG^* \!\!\!\! &\equiv & \!
\int_{0}^{\infty} \frac{dt}{t^{\nu+\epsilon}}
\sum_{m}' e^{-\pi t (m+\rho)^2-\pi \delta t} 
=
\int_{0}^{\infty} \frac{dt}{t^{\nu+\epsilon}}
\bigg[\sum_{m} e^{-\pi t(m+\rho)^2} -e^{-\pi t \rho^2}\bigg] e^{-\pi\delta t}
\label{firstg}
\nonumber\\
&=&
\int_{0}^{\infty} \frac{dt}{t^{\nu+\epsilon}}
\bigg[e^{-\pi t \Delta_\rho^2} - e^{-\pi t \rho^2}
+ \sum_{m>0} e^{-\pi t (m+\Delta_\rho)^2}+
 \sum_{m>0} e^{-\pi t (m-\Delta_\rho)^2}\bigg] e^{-\pi\delta t}
\label{secondg}\nonumber\\
&=&
\Gamma[1-\nu-\epsilon]\, \pi^{\nu-1+\epsilon}
\Big[
(\delta+\Delta_\rho^2)^{\nu-1+ \epsilon} -
(\delta+\rho^2)^{\nu-1+\epsilon}\Big]
\nonumber\\
&+&
\Gamma[1-\nu-\epsilon]\, \pi^{\nu-1+\epsilon}
\Big[
\sum_{m> 0} {[(m+\Delta_\rho)^2+\delta]^{\nu-1+\epsilon}}
+(\Delta_\rho\ra -\Delta_\rho)\Big]\nonumber\\
&=&
\!\!\!\!
\Gamma[1-\nu-\epsilon]\, \pi^{\nu-1+\epsilon}
\bigg\{\!
(\delta+\Delta_\rho^2)^{\nu-1+ \epsilon}\! -
(\delta+\rho^2)^{\nu-1+\epsilon}
+\!
\Big[\zeta[2-2\nu-2\epsilon,1+\Delta_\rho]+(\Delta_\rho\!\ra\!
-\Delta_\rho)\Big]\!\!
\bigg\}
\nonumber\\
&+&\!\!
 \pi^{\nu-1+\epsilon} \sum_{k\geq 1}
\Gamma[k+1-\nu-\epsilon]
\frac{(-\delta)^k}{k \, !} \zeta[2+2k -2\nu-2 \epsilon,1+\Delta_\rho]
+
(\Delta_\rho\!\ra\! -\Delta_\rho)\Big]\label{gstar}
\end{eqnarray}
with   $\Delta_\rho^2+\delta\not=0$. We  used the  convergent expansion  
\cite{elizalde} ($0<q/a<1$)
\begin{equation}
\sum_{m\geq 0}[a(m+c)^2+q]^{-s}=a^{-s}\sum_{k\geq 0}
\frac{\Gamma[k+s]}{k\,!\, \Gamma[s]} \bigg[\frac{-q}{a}\bigg]^k
\zeta[2 k+ 2 s, c]
\end{equation}
where $\zeta[x,c]$ with  $c\not=0,-1,-2,\cdots$ has
 one singularity (simple pole) at $x=1$
and $\zeta[x,1]=\zeta[x]$.

The above result for $\cG^*$ can be simplified for
specific cases, if $\delta\ll 1$. If $\nu$ is  such as 
$\nu= 1+N^*$  or $\nu=1/2+N^*$ with $N^*$ a non-zero
natural number, the series in $k$ has singularities 
from the $\Gamma$ and $\zeta$ functions respectively 
(note that $\epsilon\!\ra \!0$). One can isolate
such singularities from the  rest  of the series 
which can be shown to  vanish for $\delta\ll 1$. 

\noindent
For such cases  one finds for $\delta\ll 1$ (using 
$\zeta[1-\epsilon,q]=-1/\epsilon-\psi(q)+\cO(\epsilon)$, 
$\Gamma[-\epsilon]=-1/\epsilon-\gamma+\cO(\epsilon)$)
\begin{eqnarray}
\!\!\!\!\!\!\!\!\!\!\!
\cG^*\!\!\!\!\! &=&
\Gamma[1-\nu-\epsilon]\,\,  \pi^{\nu-1+\epsilon}
\Big[(\delta+\Delta_\rho^2)^{\nu-1+ \epsilon} -
(\delta+\rho^2)^{\nu-1+\epsilon}\Big]
\nonumber\\
\!\!&+&
\Gamma[1-\nu-\epsilon]\,\, \pi^{\nu-1+\epsilon}
\Big[
\zeta[2-2\nu-2 \epsilon,1+\Delta_\rho]
+(\Delta_\rho\ra -\Delta_\rho)\Big]\nonumber\\
&-&
\!\!\!\!
\delta^{K}_{N^*\!\!, \nu-{1}/{2}} 
\frac{(-\pi \delta)^{\nu-1/2}}{(\nu-1/2)\,!}
\Big[\frac{1}{\epsilon}+\psi(\Delta_\rho)+\psi(-\Delta_\rho)
+\ln(4 \pi e^\gamma)\Big]
\nonumber\\
&+&
\!\!\!
\delta_{N^*\!\!, \nu-1}^{K} \pi^{\nu-1+\epsilon}
\!\! \sum_{m\geq 2}^{\nu} 
\Gamma[m-\nu-\epsilon] \frac{(-\delta)^{m-1}}{(m-1)\,!} 
\Big[
\zeta[2 m-2 \nu-2\epsilon,1+\Delta_\rho]+
(\Delta_\rho\!\ra\! -\Delta_\rho)
\Big]\label{gstartwo}
\end{eqnarray}
where $\delta^K_{a,b}$ is a notation for  the Kronecker delta, 
equal to 1 for $a=b$ and zero otherwise, 
$N^*$  is a non-zero natural number, 
$\psi(x)\equiv d(\ln\Gamma[x])/dx$. Note that eq.(\ref{gstartwo})
has a finite number of terms only and gives a simple form for the
final result in DR of the integral in eq.(\ref{firstg}).

Further, the second integral ($\cH^*$) in eq.(\ref{qqqg}) is
\begin{eqnarray}
\cH^* & \equiv & \int_{0}^{1}\frac{dt}{t^{1/2+\nu+\epsilon}}
\Big[-1+\sqrt t \, e^{-\pi t \rho^2}\Big] e^{-\pi \delta t}
\label{qqqgg}
=
-(\pi\delta)^{\nu-1/2+\epsilon}
\Big[\Gamma[1/2-\nu-\epsilon]-\Gamma[1/2-\nu-\epsilon,\pi\delta]\Big]
\nonumber\\
&+&
[\pi(\delta+\rho^2)]^{\nu+\epsilon-1}\Big[\Gamma[1-\nu-\epsilon]-
\Gamma[1-\nu-\epsilon,\pi(\delta+\rho^2)\Big]
\end{eqnarray}
The general result for $\cR^*$ of eq.(\ref{ixig}) is, with
$\Delta_\rho^2+\delta\not=0$, $\xi\ll 1$, $\delta\ll 1$
\begin{equation}
\cR^*=\cI_\xi^*+\cF^*=\cI_\xi^*+\cG^*+\cH^*
\end{equation}
with $\cI_\xi$ given in eq.(\ref{ixiig}), $\cF^*$ in eq.(\ref{qqqg}),
$\cG^*$ in eq.(\ref{gstartwo}) (or (\ref{gstar})) and 
$\cH^*$ in eq.(\ref{qqqgg}). 
Additional assumptions are needed to simplify this result further.

As an example, if $\nu=3/2$ one has 
\begin{eqnarray}\label{R2*_final}
\!\!\!\!\!\!\!\!\!\!
\cR^*[\xi,\rho,\delta,{3}/{2}]\!\!\!
&\equiv& \int_{\xi}^{\infty} \frac{dt}{t^{3/2}} \sum_{m}'
e^{-\pi t (m+\rho)^2 -\pi \delta t} \nonumber\\
&=& 
\frac{1}{\xi}-\frac{2}{\sqrt \xi} e^{-\pi(\delta+ \rho^2) \xi}
-
2\pi (\delta+\rho^2)^{\frac{1}{2}} \,
\textrm{Erf}[(\pi\xi(\delta+\rho^2))^{\frac{1}{2}}]\nonumber\\
&+&
2\pi\Big[(\delta+\rho^2)^{\frac{1}{2}}+
\frac{1}{6}+
\Delta_{\rho}^2-(\Delta_{\rho}^2+\delta)^{\frac{1}{2}}\Big]+
\pi\delta\ln\Big[4\pi \, \xi\,
e^{\gamma+\psi(\Delta_\rho)+\psi(-\Delta_\rho)}\Big], \label{exam}
\end{eqnarray}
with $\Delta_\rho^2+\delta\not=0$. One can set  $\delta=0$
 to  obtain $\cR_2$ of eq.(\ref{r2ap}).
If also $\xi(\rho^2+\delta)\ll 1$ then the term proportional to Erf function 
is also absent.  $\cG^*$ for $\nu=3/2$ is the  DR version of
 (\ref{exam}) and has a similar form, with  the above
 $\xi$ dependence  replaced by $\pi\delta/\epsilon$.
%%%%  For $\nu=1$ eq.(\ref{a12}) is also recovered. 

The method presented  is particularly useful for cases with
$\nu=N^*+1/2$. The methods also provides a dimensional 
regularisation (DR)  version of the initial integral $\cR^*$, 
given by $\cG^*$ of eq.(\ref{gstartwo}),
and thus a general relation between series of integrals computed  in the
DR and proper-time cutoff regularisation schemes.

%%%%%%%%%%%%%%%%%%%%%%%%%%%%%%%%%%%%%%%%%%%%%%%%%%%%%%%%%%%%%%%%%
\newpage
\section{Case 2 (B) for  $\rho_2 \not \in \bZ$ and $\rho_1\in \bZ$.}
\setcounter{equation}{0}
\label{case3}

%%%%%%%%%%%%%%%%%%%%%%%%%%%%%%%%%%%%%%%%%%%%%%%%%%%%%%%%%%%%%%%%%

We  extend the validity of Case 2 (B) in the text to situations 
when $\rho_2$ is non-integer and $\rho_1$ integer. The method of
Case 2 (B) is  not well-defined for such a case, 
see integral $\cL_2$ eq.(\ref{cL123}) 
which is IR divergent if $\rho_1\in\bZ$. However the
(formal)  limit of the final result of 
Case 2 (B) for $\rho_2$ non-integer, $\rho_1$ integer 
is finite and does give the correct result as we show below
by computing separately this case. 
This finding  provides an extension of Case 2 (B) to all cases
with $\rho_1$ {\it or} $\rho_2$  non-integer with arbitrary, 
real values for  the other. The correction  $\Omega_i$ can be written as 
\begin{equation}
\Omega_i=\frac{\beta_i}{4\pi} \cJ_0^{(4)}+
\frac{\overline \beta_i}{4\pi}\cJ^{(4)}
\end{equation}
where we introduced:
\begin{eqnarray}
\!\!\!\!\!\!\!\cJ_0^{(4)}& \equiv &\int_{\xi}^{\infty} \frac{dt}{t} 
e^{-\pi\,t \,M^2_{0,0}/\nu^2}=\int_{\xi}^{\infty} \frac{dt}{t} 
e^{-\pi \, t\, \xi \vert \rho_2 - U \rho_1 \vert^2/(T_2 U_2)}=
\Gamma[0,\pi\xi \vert\rho_2-U \rho_1\vert^2/(T_2 U_2)]\nonumber\\
\nonumber\\
\!\!\!\!\!\!\!
\cJ^{(4)} & \equiv & \sum_{m_1,m_2}' \int_{\xi}^{\infty} \frac{dt}{t}
e^{-\pi \, t\, M^2_{m_1,m_2}/\mu^2} 
=\!\!\! \sum_{m_1,m_2}' \int_{\xi}^{\infty} 
\frac{dt}{t} 
e^{-\frac{\pi \, t}{T_2 U_2} \vert m_2+
\rho_2-U (m_1+\rho_1)\vert^2} \nonumber\\
\nonumber\\
&=&\!\!\!\!\!\! \sum_{m,m_2}' \int_{\xi}^{\infty}\frac{dt}{t} 
e^{-\frac{\pi\, t}{T_2 U_2} \vert m_2+\rho_2 -U m\vert^2}
\!\!\!-\Gamma[0,\pi \xi \vert \rho_2 -U \rho_1\vert^2/(T_2 U_2)]
+\Gamma[0,\pi \xi \rho_2^2/(T_2 U_2)]
\end{eqnarray}
In the last step, under the integral, the exponential
evaluated for $(m_1,m_2)=(0,0)$ was added and subtracted, 
we then replaced  $m_1+\rho_1\ra m$ as the new summation index,  
and finally isolated  the ``new'' $(0,0)$ mode  from the rest  of the series.
Further, the integrand in the final   series can 
be written after Poisson re-summation as
\begin{eqnarray}
&&\!\!\!\!\!\!\!\!\!\!\!\!
\sum_{m,m_2}' e^{-\frac{\pi t}{T_2 U_2} \vert m_2+\rho_2-U m\vert^2} =
\sum_{m_2}'e^{-\frac{\pi \, t}{T_2 U_2} (m_2+\rho_2)^2}
+\sum_{m}' \sum_{m_2} 
e^{-\frac{\pi \, t}{T_2 U_2} \vert m_2+\rho_2-U m\vert^2}\nonumber\\
\nonumber\\
&=& 
\sum_{m_2}'e^{-\frac{\pi \, t}{T_2 U_2} (m_2+\rho_2)^2}+
\bigg[\frac{T_2 U_2}{t}\bigg]^{\frac{1}{2}}\bigg[
\sum_{m}'  e^{-\pi \, t\, m^2 \frac{U_2}{T_2}}
+ 
\sum_{m}' \sum_{\tilde m_2}' 
e^{-\pi \, \tilde m_2^2 \frac{T_2 U_2}{t} 
-\pi\, t\, m^2  \frac{U_2}{T_2} +2 i \pi \tilde m_2 (\rho_2-U_1 m)}
\bigg]\nonumber
\end{eqnarray}
Since $\rho_2$ is assumed non-integer, each of the above
series can be integrated separately over $(\xi,\infty)$ to 
compute $\cJ^{(4)}$. Further
\begin{eqnarray}
\cK_1
&\equiv &
\int_{\xi}^{\infty}\frac{dt}{t} \sum_{m_2}' e^{-\frac{\pi \, t}{T_2
U_2} (m_2+\rho_2)^2}=2 \bigg[\frac{T_2 U_2}{t}\bigg]^{\frac{1}{2}}
-\Gamma[0,\pi \xi \rho_2^2/(T_2 U_2)]-
\ln \vert 2 \sin \pi \rho_2\vert^2\nonumber\\
\nonumber\\
\cK_2& \equiv & (T_2 U_2)^{\frac{1}{2}}
\int_{\xi}^{\infty} \frac{dt}{t^{3/2}} \sum_{m}'
e^{-\pi\, t \, \frac{U_2}{T_2} m^2}
=\frac{T_2}{\xi}-2 \bigg[\frac{T_2 U_2}{\xi}\bigg]^{\frac{1}{2}}
+\frac{\pi}{3} U_2 \nonumber\\
\nonumber\\
\cK_3 & \equiv  & (T_2 U_2)^{\frac{1}{2}}
\int_{\xi}^{\infty}\frac{dt}{t^{3/2}}
\sum_{m}'\sum_{\tilde m_2}' e^{- \pi \, \tilde m_2^2 \frac{T_2 U_2}{t} 
-\pi\, t\, m^2  \frac{U_2}{T_2} +2 i \pi \tilde m_2 
(\rho_2-U_1 m)}\nonumber\\
\nonumber\\
&=& -\ln\prod_{m\geq 1}
\Big\vert 1-e^{2 \,i \pi m U} \, e^{-2 i \pi  \rho_2}
\Big\vert^2 \, \Big\vert 1- e^{2 i \pi m U} \, 
e^{2 i \pi \rho_2 }\Big\vert^2
\end{eqnarray}
which are valid provided that  $T_2 U_2/\xi\ll\max \{1/U_2^2,U_2^2\}$.
To evaluate $\cK_1$ we used eq.(\ref{limit1}) while to evaluate
$\cK_2$ we  used eq.(A-12) in Appendix A of ref.\cite{Ghilencea:2002ff}
(which agrees with the limit of $\rho\!\ra\! 0$ in
eq.(\ref{R2_final})). $\cK_3$ can be evaluated in the limit 
$\xi\ll 1$ using eq.(\ref{bessel1}).

Adding together all the contributions $\cK_i$ one finds for $\Omega_i$ 
\begin{eqnarray}\label{omega_c3} 
\Omega_i & = & 
\frac{\beta_i}{4\pi}\ln \frac{\Lambda^2 (R_2 \sin\theta)^2}{\pi e^\gamma
\vert \rho_2 - U \rho_1\vert^2}
- \frac{{\overline {\beta}}_i}{4 \pi} 
\ln\Big[4 \pi e^{-\gamma} \, e^{-{T_2^*}} \, {T_2^*} \, U_2
\, \vert \eta(U)\vert^4\Big]
-\frac{\overline \beta_i}{4\pi} \ln \bigg\vert 
\frac{\sin \pi \rho_2}{\pi (\rho_2- U
\rho_1)}\bigg\vert^2\nonumber\\
\nonumber\\
&-&\frac{\overline \beta_i}{4\pi}
\ln\bigg[\prod_{m\geq 1}\Big \vert 1- e^{2 i \pi m U}\Big\vert^{-4}
\Big\vert 1-e^{2 i \pi m U} \, e^{-2 i \pi  \rho_2}
\Big\vert^2 \, \Big\vert 1- e^{2 i \pi m U} \, 
e^{2 i \pi \rho_2 }\Big\vert^2\bigg]
\end{eqnarray}
The first term in $\Omega_i$ is the contribution of the original 
$(0,0)$ modes. The second contribution is due to the tower of
Kaluza-Klein modes of non-zero level. The third term 
(divergent in the limit $\rho_2$ integer)  bears some similarities 
with the third term in eq.(\ref{w3}) of the one-dimensional case.
The last term above is suppressed for large $U_2$ and is a
two-dimensional effect.  An equivalent form of the above result is
\begin{eqnarray}\label{omega_c3f}
\Omega_i=\frac{\beta_i}{4\pi} \ln \frac{\Lambda^2 (R_2
\sin\theta)^2}{\pi e^\gamma \vert \rho_2 -U \rho_1\vert^2}
+
\frac{\overline \beta_i}{4\pi} \bigg[T_2^* 
-\ln \frac{T_2^* U_2}{\pi e^\gamma \vert \rho_2 -U \rho_1\vert^2}
-
\ln\bigg\vert \frac{\vartheta_1 (\Delta\rho_2\vert U)}
{\eta(U)}\bigg\vert^2
\bigg]
\end{eqnarray}
where the special function $\vartheta_1$ was defined  in (\ref{R3_3})
and $T_2^*=T_2/\xi$.
The result (\ref{omega_c3f}) agrees with the formal limit $\rho_1$
integer ($\rho_2$ non-integer) of  eq.(\ref{equiv_form}) of Case 2  
(B) in the text. The analysis of Case 2 (B) is then  valid  as long as 
$\rho_1$ {\it or } $\rho_2$ is non-integer with arbitrary, real values
for  the other.

\end{document}